\documentclass[sigconf]{acmart}
\AtBeginDocument{%
  }

\usepackage{bbold}
\setcopyright{acmlicensed}
\copyrightyear{2025}
\acmYear{2025}
\acmDOI{XXXXXXX.XXXXXXX}
\acmConference[Mobihoc '25]{The 26th International Symposium on Theory, Algorithmic Foundations, and Protocol Design for Mobile Networks and Mobile Computing}{October 27-30, 2025}{Houston, USA}
\acmISBN{979-8-4007-XXXX-X/2025/10}
\input{mysymbol.sty}

\usepackage{algorithm}
\usepackage{algpseudocode}
\usepackage{subfig}
\usepackage{enumitem}
\usepackage{multirow}

\setlist[itemize]{topsep=0pt, itemsep=0pt, parsep=0pt}
\setlist[enumerate]{topsep=0pt, itemsep=0pt, parsep=0pt}

\begin{document}

\title{SeLR: Sparsity-enhanced Lagrangian Relaxation for Computation Offloading at the Edge}

% \begin{verbatim}
% \title[Sparse Dual Lagrangian Methods for Convex Optimization in computation offloading]{Integrating Sparsity and Dual Lagrangian Methods for Convex Optimization in computation offloading within Communication Networks}
% \end{verbatim}

\author{Negar Erfaniantaghvayi}
\affiliation{%
  \institution{Rice University}
  \city{Houston}
  \state{Texas}
  \country{USA}
}
\email{ne12@rice.edu}

\author{Zhongyuan Zhao}
\affiliation{%
  \institution{Rice University}
  \city{Houston}
  \state{Texas}
  \country{USA}
}
\email{zhongyuan.zhao@rice.edu}

\author{Kevin Chan}
\affiliation{%
  \institution{DEVCOM Army Research Laboratory}
  \city{Adelphi}
  \state{MD}
  \country{USA}
}
\email{kevin.s.chan.civ@army.mil}

% \author{Gunjan Verma}
% \affiliation{%
%   \institution{DEVCOM Army Research Laboratory}
%   \city{Adelphi}
%   \state{MD}
% \email{gunjan.verma.civ@army.mil}

\author{Ananthram Swami}
\affiliation{%
  \institution{DEVCOM Army Research Laboratory}
  \city{Adelphi}
  \state{MD}
  \country{USA}
}
\email{ananthram.swami.civ@army.mil}

\author{Santiago Segarra}
\affiliation{%
  \institution{Rice University}
  \city{Houston}
  \state{Texas}
  \country{USA}
}
\email{segarra@rice.edu}

\renewcommand{\shortauthors}{Erfaniantaghvayi et al.}

% The framework guarantees that tasks are offloaded to destinations with sufficient computational resources (e.g., CPU, RAM) to meet workflow requirements while adhering to bandwidth constraints. 

\begin{abstract}
This paper introduces a novel computational approach for offloading sensor data processing tasks to servers in edge networks for better accuracy and makespan. 
A task is assigned with one of several offloading options, each comprises a server, a route for uploading data to the server, and a service profile that specifies the  performance and resource consumption at the server and in the network.
This offline offloading and routing problem is formulated as mixed integer programming (MIP), which is non-convex and HP-hard due to the discrete decision variables associated to the offloading options.
The novelty of our approach is to transform this non-convex problem into iterative convex optimization by relaxing integer decision variables into continuous space,  
combining primal-dual optimization for penalizing constraint violations and reweighted $L_1$-minimization for promoting solution sparsity, which achieves better convergence through a smoother path in a continuous search space. 
Compared to existing greedy heuristics, our approach can achieve a better Pareto frontier in accuracy and latency, scales better to larger problem instances, and can achieve a 7.72--9.17$\times$ reduction in  computational overhead of scheduling compared to the optimal solver in hierarchically organized edge networks with 300 nodes and 50--100 tasks.
\end{abstract}

\begin{CCSXML}
<ccs2012>
   <concept>
       <concept_id>10003033.10003099.10003100</concept_id>
       <concept_desc>Networks~Cloud computing</concept_desc>
       <concept_significance>300</concept_significance>
       </concept>
   <concept>
       <concept_id>10003033.10003068.10003073.10003074</concept_id>
       <concept_desc>Networks~Network resources allocation</concept_desc>
       <concept_significance>500</concept_significance>
       </concept>
   <concept>
       <concept_id>10002950.10003624.10003625.10003630</concept_id>
       <concept_desc>Mathematics of computing~Combinatorial optimization</concept_desc>
       <concept_significance>500</concept_significance>
       </concept>
   <concept>
       <concept_id>10003752.10003809.10003716.10011141.10010045</concept_id>
       <concept_desc>Theory of computation~Integer programming</concept_desc>
       <concept_significance>500</concept_significance>
       </concept>
   <concept>
       <concept_id>10003752.10003809.10003716.10011138.10010041</concept_id>
       <concept_desc>Theory of computation~Linear programming</concept_desc>
       <concept_significance>500</concept_significance>
       </concept>
   <concept>
       <concept_id>10003752.10003809.10003636.10003808</concept_id>
       <concept_desc>Theory of computation~Scheduling algorithms</concept_desc>
       <concept_significance>500</concept_significance>
       </concept>
 </ccs2012>
\end{CCSXML}

\ccsdesc[300]{Networks~Cloud computing}
\ccsdesc[500]{Networks~Network resources allocation}
\ccsdesc[500]{Mathematics of computing~Combinatorial optimization}
\ccsdesc[500]{Theory of computation~Integer programming}
\ccsdesc[500]{Theory of computation~Linear programming}
\ccsdesc[500]{Theory of computation~Scheduling algorithms}
\keywords{Edge computing, multi-hop networks,  reweighted $L_1$-minimization, primal-dual optimization, resource allocation, mixed-integer programming.}


\maketitle

\section{Introduction}
% \textcolor{red}{SS: Please make sure to add more references to our work. For sure, you should add references to Zhongyuan's ICASSP 2024 and ICASSP 2025 papers on computation offloading. You might also have an opportunity to cite the work on routing? (your previous paper and a few of Zhongyuan's too).}
To ensure timely processing of sensory data, computational resources can be placed close to data-generating sensors at the network edge to reduce latency and congestion in communications.
This edge computing paradigm is critical to applications based on real-time data analytics, such as video surveillance and environmental monitoring, where the popularity of Internet of Things (IoT) sensors, resource-intensive machine learning techniques, and increasing demands for low-latency, high-resolution services have been driving the advancements in task offloading and network resource allocation~\cite{ullah2023optimizing, patsias2023task, dinh2017offloading, xiao2020edgeabc, tran2018joint, jiang2022joint, kan2018task, yan2019optimal, yang2023task}. 
In edge networks, sensors and computing servers are connected with wired and/or wireless links, potentially over multiple hops.
Unlike in cloud facility where servers are interconnected via high-speed fiber networks, computation offloading in edge networks is often constrained by the link capacity and network topology~\cite{guo2023intelligent, yan2019optimal}.
The constraints in network connectivity, together with diverse task profiles in resource consumption and performance requirements, require joint decision-making for offloading and routing, presenting a unique challenge for task scheduling in edge networks. 
% Emerging applications in areas such as real-time analytics, edge computing, machine learning pipelines, and the Internet of Things (IoT) require network systems to process a massive influx of computational tasks efficiently \cite{xiao2020edgeabc, tran2018joint, jiang2022joint, kan2018task, yan2019optimal, yang2023task}. 
% These tasks often vary widely in terms of computational complexity, accuracy requirements, and latency sensitivity. 
% The challenge lies in ensuring that these tasks are allocated to the most appropriate computational resources while meeting stringent performance metrics and efficiently utilizing the network bandwidth \cite{guo2023intelligent, yan2019optimal}.

Currently, methods of task offloading in edge networks can be categorized as online and offline approaches. 
Online task offloading schedules tasks asynchronously in a distributed manner~\cite{Kamran2022deco,lin2020distributed,jiang2023joint,liu2019dynamic,Bi2021Lyapunov,zhao2025icassp, erfaniantaghvayi2024ant}, making offloading decisions as soon as they are initialized. 
However, online offloading mostly employ separated offloading and routing decision-making based on limited local information~\cite{Kamran2022deco,lin2020distributed,jiang2023joint,liu2019dynamic,Bi2021Lyapunov}, trading off optimality for real-time adaptivity.
Distributed scheme for joint offloading and routing~\cite{zhao2025icassp} has been recently developed by modeling computing as sending packets to a virtual sink over a virtual link, transforming joint offloading and routing into a routing problem. 
However, this approach is less flexible in performance-cost trade offs and may not be feasible for applications without full control of low-level networking protocols.
Offline offloading typically utilizes a centralized scheduler to process tasks in batches, in which a batch of offloading and routing decisions is formulated as a MIP problem~\cite{Müller2015computation,liu2020distributed,kiamari2022gcnscheduler}.
The MIP formulation provides better flexibility for trade-offs in task performance and resource cost, and is suitable for tasks with larger data sizes and lower arrival rates in small to medium-sized networks. 
In addition, distributed offline offloading based on graph neural networks (GNNs) has also been developed to balance optimality and scalability~\cite{zhao2024congestionaware}.

In this paper, we focus on offline joint task offloading and routing problems, formulated as MIP with an objective of maximizing the total utility of task assignments subject to various constraints~\cite{gerogiannis2022deep, bi2020joint, xiao2020parameter, wang2019edge, ning2020intelligent,Müller2015computation,liu2020distributed,kiamari2022gcnscheduler}. 
The utility function of an offloading option can be based on a single performance metric, such as makespan~\cite{Müller2015computation,liu2020distributed,kiamari2022gcnscheduler} or a multiple metrics such as accuracy, latency, and energy efficiency~\cite{zhao2020improving,younis2024energy}.
The constraints often include limits in computational resources, e.g., CPU and memory capacities, limits on communication bandwidth, and minimal requirements on performances such as makespan, resolution, and accuracy. 
Such decision problems are often non-convex and NP-hard due to their discrete nature, rendering exact solvers impractical for real-time applications and/or in large networks~\cite{lee2011mixed, belotti2013mixed, burer2012non, floudas1995nonlinear, tawarmalani2013convexification}, whereas fast heuristics like greedy search often have large optimality gaps.

To balance optimality and computational complexity, we develop an efficient approach for offline task offloading in edge networks by integrating two iterative methods, reweighted $L_1$-minimization~\cite{candes2008enhancing} and primal-dual optimization~\cite{kochenderfer2019algorithms}.
Our approach transforms the MIP problem into convex optimization by relaxing discrete decision variables into continuous space, while iteratively updating the Lagrangian multipliers and $L_1$ weights to encourage sparsity in solutions. 
This iterative convex relaxation can substantially reduce the computational burden compared to exact MIP solvers. 
% While relaxation-based methods are widely used in optimization, our contribution lies in tailoring and integrating such techniques into a scalable and iterative framework specifically designed for edge networks with tight performance and resource constraints.
Moreover, we design an algorithmic framework to further improve the efficiency by warm starting the primal-dual optimization, and partially assigning tasks with integer solutions in the iterations to progressively reduce the problem scale.

Our problem formulation enables joint optimization of task placement and routing under various constraints in server capacity, communication bandwidth, and network topology, with the goal of maximizing execution accuracy and minimizing latency. 
With numerical experiments on networks of 300 nodes, our approach is demonstrated to achieve better scalability and a better Pareto frontier in accuracy and latency, comparable to multiple greedy approaches, while offering a 7--9 times reduction in computational overhead for scheduling 50--100 tasks in hierarchically organized edge networks with 300 nodes. 
% We evaluate our approach across a range of network topologies and task profiles, benchmarking it against established offloading schemes. 
% Experimental results demonstrate that our method consistently delivers superior trade-offs between accuracy, latency, and runtime, making it a practical and adaptable solution for computation offloading in complex edge network environments.

% \vspace{1mm}

\noindent 
\textbf{Contribution.} The contributions of this paper are threefold:
% \textcolor{red}{SS: Related to the comment above, maybe combine points (1) and (2). Point (1) right now it literally says relaxing integer variables which is a pretty standard procedure. That by itself is not a contribution.}:
\begin{enumerate}[leftmargin=1.5em]
    % \item We propose a computation offloading framework that integrates task-specific requirements (e.g., latency and accuracy) with network-level constraints (e.g., computational resources, bandwidth, and hop count) to optimize task allocation in communication networks.
    \item We introduce Sparsity-enhanced Lagrangian Relaxation (SeLR), a novel iterative approach that combines primal-dual optimization for penalizing constraint violations, and reweighted $L_1$-minimization for encouraging sparsity, which can efficiently approximate integer solutions of a non-convex problem within a convex framework. 
    % This integration enables the use of computationally efficient solvers while maintaining scalability and practical feasibility.
    \item We further improve the convergence speed by developing an algorithm that warm starts SeRL with a valid solution and progressively reduces the problem size by removing partial solutions from decision variables.
    % \item We convert the non-convex optimization problem into a convex formulation by relaxing integer decision variables to continuous ones, allowing the use of computationally efficient convex solvers.
    % \item We propose a novel approach that integrates reweighted $L_1$-normalization with the dual Lagrangian method in our convex formulation, effectively approximating integer solutions while ensuring practical feasibility and scalability.
    \item Through numerical experiments, we demonstrate that our approach can achieve better Pareto frontiers in accuracy and latency compared to other heuristic approaches, with a substantially lower runtime compared to the optimal solver.
\end{enumerate}

\begin{figure}[!t]
    \includegraphics[width=1\linewidth]{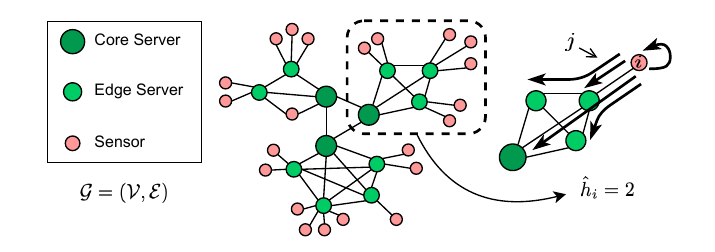}
    \vspace{-0.1in}
    \caption{Illustration of the simplified network topology, highlighting the connections between servers and sensors in our setup. Offloading Options for each task are selected based on the network topology, taking into account the max-hop limitation, edge bandwidth, and the tasks' accuracy and latency requirements, as outlined in Algorithm~\ref{alg:rworkflow_construction}.
    }
    \label{fig:main}
    \vspace{-0.1in}
\end{figure}

\section{System model}\label{sec:system}
We model a multi-hop network, as exemplified in Fig.~\ref{fig:main}, as a directed connectivity graph $\mathcal{G} = (\mathcal{V}, \mathcal{E})$, 
% \textcolor{red}{SS: I understand that the kind of constraints that we use in the optimization problem are more applicable to the wired case, but I consider that as a simplification of the optimization problem rather than the system model. Maybe here avoid the word `wired' as typically the kind of edge networks that people might have in mind (such as IoT) are typically wireless} 
where a node $v\in\mathcal{V}$ represents a sensor or a server in the network, and an edge $e = (v_a, v_b) \in\mathcal{E}$ indicates that node $v_a$ can directly transmit data to node $v_b$ via a wired or wireless link. 
We assume $\mathcal{G}$ to be strongly connected, i.e., any two nodes within the network can reach each other via a directed path, and that in general $(v_b,v_a)\in\ccalE$ if $(v_a,v_b)\in\ccalE$.
The node set $\ccalV$ can be split into a sensor set $\ccalS$, and a server set $\ccalW$, where $\ccalS\cup\ccalW=\ccalV$, $\ccalS\cap\ccalW=\emptyset$.
Each node $v \in \mathcal{V}$ is associated with a set of attributes, such as device-type $s_v=\mathbb{1}(v\in\ccalW)$, CPU, GPU, and RAM capacities, denoted as $C^{\text{CPU}}_{v}, C^{\text{GPU}}_{v}, C^{\text{RAM}}_{v} $. 
A sensor node $v\in\ccalS$, lacks GPUs and has lower CPU and RAM capacities, whereas a server node $v\in\ccalW$ has richer computational resources. 
% Nodes designated as sensors lack GPUs and have lower CPU and RAM resources compared to other device types, such as edge servers and core servers.
The bandwidth capacity of a link $e \in \mathcal{E}$ is denoted as $C^{\text{BW}}_e$, and the capacities of links with opposite directions could be different. 
% A simplified representation of our network topology is shown in Figure~\ref{fig:main}.
% \textcolor{red}{SS: Point to a figure of the network topology here.}

We consider a set of homogeneous tasks, denoted as $\ccalT$, each originates from a unique sensor, and is identified by its source, e.g., $\ccalT \subseteq \ccalS$.  
% \textcolor{red}{SS: Strictly speaking, N will never be equal to the cardinality of $\mathcal{V}$, right? since you have as many tasks as sensor nodes and the cardinality of $\mathcal{V}$ contains all node types}
% In this work, for simplicity, we assume that no two tasks can originate from the same node.
A task $i\in\ccalT$ can be offloaded to servers with richer computational resources for improved execution quality (e.g., accuracy) and makespan (total latency), or executed on-device with reduced quality and/or increased makespan if none of the servers has sufficient resources. 
The maximum allowable latency and minimal required accuracy for task $i$ are denoted as $\hat{\tau}_i$ and $\hat{q}_i$, respectively.
We can further limit offloading options for task $i$ to nodes its $\hat{h}_i$-hop neighborhood. 
% Additionally, the \textit{offloading options} of a task can be limited to a prescribed $h$-hop neighborhood of the source sensor.

A server or sensor node $v\in\ccalV$ may have multiple \textit{service profiles} that define a set of trade-offs for task execution cost and performance, denoted as $\ccalP_v$.
Each service profile $p\in\ccalP_v$ comprises a set of attributes such as \textit{accuracy} $q_p$, \textit{latency} $\tau_p$, \textit{RAM cost} $c^{\text{RAM}}_{p}$, \textit{CPU cost} $c^{\text{CPU}}_{p}$, \textit{GPU availability}, \textit{bandwidth cost} $c^{\text{BW}}_{p}$, \textit{max-hop} $h_p$, and \textit{device-type} $s_p$.
The device-type indicates whether the node is a sensor for on-device execution or a server.
Max-hop $h_p$ limits the acceptable tasks to those originated from sensors within the $h_p$-hop neighborhood of the node (including itself), denoted as $\ccalN^+(v;h_p)$.  
For a sensor, we can set $h_p=0$ to prevent it from serving tasks from other sensors.
Accuracy and latency help determine whether the service profile meets the minimal performance requirement of a task. 
RAM, CPU, and bandwidth costs indicate the computational resources the service will consume.  
GPU availability impact the latency and quality of task execution.

% \textcolor{red}{SS: I think that this thing of workflow and resource workflow can still be confusing. For the workflows, the idea is that we have measured accuracies and latencies, so that if you tell me device-type, number of hops, RAM, CPU, and bandwidth, we have established the accuracy and latency achievable in this setting. And we then see the requirements of the task and the nodes near the corresponding sensor node and consider as candidate workflows those that satisfy the minimum accuracy and maximum latency permissibe for the task. E.g., the rightmost part of Figure 1 where you compare the accuracy of the workflow with that of the task I think it is useful to understand that, but you do not explain it in the text. You can expand that description here a bit. I saw later that this is explained more in Algorithm 2 after ... maybe you can mention that too.}

Based on the network topology $\ccalG$, service profiles of each node, and performance requirements of each task, we can create a set of valid \textit{offloading options} for each task $i\in\ccalT$, denoted as $\ccalJ_i$, using Algorithm~\ref{alg:rworkflow_construction}.
Each \textit{offloading option} $j=(t_j, v_j, r_j, p_j)\in\ccalJ_i$ is a unique combination of task $t_j$, server $v_j$, route $r_j$, and service profile $p_j$, of which the profile of accuracy $q_j$ and latency $\tau_j$, as well as CPU, RAM and bandwidth costs are inherited from the service profile $p_j$, e.g., $q_j=q_{p_j}$, as shown in lines 6-7 in Algorithm~\ref{alg:rworkflow_construction}.
Each task will be assigned with one and only one offloading option.
It is important to note that the number of valid options $|\ccalJ_i|$ for task $i\in\ccalT$ depends on its performance requirement and network locality, and can vary among tasks.
% \textcolor{red}{You are no longer using the UGS term} 
In the exemplary network in Fig.~\ref{fig:main}, a task $i$ has five offloading options that meet its performance requirement and max-hop $h_{p}=2$. 
In addition, we define the set of all offloading options across the network as $\ccalJ=\bigcup_{i\in\ccalS}\ccalJ_i$.
% Among them, $rw_{t_1}^3$ results in on-device task execution, while the others offload the task to servers located within a maximum of two hops from sensor originating task $t_1$.

% When assigning service profiles to tasks, after finding all nodes within the maximum hop limits of the task's origin and the associated routs, all service profiles' attributes are evaluated to identify the \textit{offloading option}. During this process, each service profile is assigned a path from the task's origin to a node within the hop limit that matches the service profile's \textit{device-type}. 
% Workflows with assigned paths are referred to as \textit{offloading options}.

% \textcolor{red}{SS: I don't understand the meaning of the next sentence ... why is this notable?} Notably, this assignment does not exhaust computational resources such as edge bandwidth and server processing power. 
% \textcolor{red}{SS: Why are we repeating this here?} As previously mentioned, the number of offloading options $|\mathcal{J}_{t_i}|$ is not necessarily the same for all tasks.

\section{Task Allocation Problem Formulation}\label{sec:problem}

To optimize both performance and resource utilization, we adopt a MIP formulation for the optimal computation offloading assignment in~\eqref{eq:mip}, by denoting the decision variable of selection an offloading option $j$ as $x_j\in\{0,1\}$, with vector $\bbx=[x_j\mid j\in\ccalJ]$ collects the decision variables for all offloading options, and a utility function for each offloading option that balances the accuracy and latency: 
% Specifically, the objective is to assign $|\ccalT|$ tasks to offloading options, ensuring that each task is assigned exactly one offloading option. The goal is to balance the trade-off between accuracy and latency.
%The goal is to maximize the accuracies and minimize the latencies across all offloading options while ensuring that all $N$ tasks are successfully allocated throughout the network. 
% This task allocation problem is formulated as the following non-convex integer optimization problem:
% \textcolor{red}{SS: In objective function below, the set J should have $t_i$ as the subindex, right? This is true all over the place, you sometimes use $i$ and sometimes $t_i$. Please make it consistent. Also, the accuracy and latency should be function of $rw_{t_i}^j$ instead of $j$, right?}
\begin{subequations} \label{eq:mip}
\begin{align}
\bbx^*=&\argmin_{\bbx} \sum_{j \in \mathcal{J}}  x_{j}\overbrace{[\lambda \cdot \tau_{j}-(1-\lambda) \cdot q_{j}]}^{u_{j}} , \label{eq:mip:obj}\\
% \text{s.t.  }  u_{j} &=  \lambda \cdot \tau_{j}-(1-\lambda) \cdot q_{j}, \forall\; j\in\ccalJ \label{eq:mip:utility} \\
\text{s.t.  }\;\;\;C^{\text{CPU}}_v &\geq  \sum_{j \in \ccalJ}  x_{j} c^{\text{CPU}}_{j,v}, \; \forall v \in \mathcal{V},\label{eq:mip:cpu} \\
C^{\text{RAM}}_v &\geq \sum_{j \in \ccalJ} x_{j} c^{\text{RAM}}_{j,v}, \; \forall v \in \mathcal{V},\label{eq:mip:ram}\\ 
C^{\text{BW}}_e &\geq \sum_{j \in \ccalJ} x_{j} c^{\text{BW}}_{j,e}, \; \forall e \in \mathcal{E}, \label{eq:mip:bw}\\
C^{\text{GBW}} &\geq   \sum_{e \in \mathcal{E}} \sum_{j \in \mathcal{J}} x_{j} c^{\text{BW}}_{j,e}  , \label{eq:mip:gbw}\\
1 &= \sum_{j \in \ccalJ_i} x_{j}, \; \forall i \in \ccalT, \label{eq:task_assignment}\\ 
x_{j} &\in \{0, 1\}, \; \forall \; j \in \ccalJ,\; \bbx=[x_j|j\in\ccalJ]\;.\label{eq:mip:binary}
\end{align}
\end{subequations}
The constraints in~\eqref{eq:mip} are explained as follows:
Constraints \eqref{eq:mip:cpu} and \eqref{eq:mip:ram} specify that the CPU and RAM consumption on a node $v$ are limited by the corresponding capacities, and \eqref{eq:mip:bw} specify that the bandwidth consumption on a link should not exceed its capacity.
Notice that constraint \eqref{eq:mip:bw} does not consider the interaction of different flows going through the same link. 
In many cases, especially in wireless links, interference effects might lead to underutilization of the edge bandwidth. 
To account for this, in \eqref{eq:mip:gbw}, we set a limit on the maximum global  bandwidth as the network capacity $C^{\text{GBW}}$.
Here, $c^{\text{RAM}}_{j,v}=c^{\text{RAM}}_{j}\mathbb{1}(v= v_j)$, $ c^{\text{CPU}}_{j,v}=c^{\text{CPU}}_{j}\mathbb{1}(v= v_j)$, $ c^{\text{BW}}_{j,e}=c^{\text{BW}}_{j}\mathbb{1}(e\in r_j)$, capture the corresponding costs of offloading option $j$ on node $v$ and link $e$, e.g., if $v$ is not the destination of $j$, $v_j$, or link $e$ is not on the route of option $j$, then the cost is zero. 
Constraint \eqref{eq:task_assignment} specifies that each task $i\in\ccalT$ must be assigned with one and only one offloading option.
\eqref{eq:mip:binary} defines the decision variable $x_j$ as binary, and vector $\bbx$ collects the decision variables for all offloading options across the network.

\begin{algorithm}[!t]
\caption{Forming Tasks' Offloading Options}
\label{alg:rworkflow_construction}
\begin{algorithmic}[1]
\Require Task $i$ originating from sensor node $v\in\ccalS$
\Ensure Output the set of offloading options for task $i$, $\ccalJ_i$

\State $\ccalJ_i=\emptyset$; extract task requirements $\hat{h}_i,\hat{\tau}_i,\hat{q}_i $
% \State Extract all available nodes within  $h$-hop of node $v$, $\ccalN^+_{h}(v)$.

% \State \textbf{Filter and Validate Service Profiles:}
\For{ node $d \in \ccalN^+(v;\hat{h}_i)$}
    \State Find all routes from $v$ to $d$ as set $\rho^d_v$
    \For{Each route $r\in \rho^d_v$}
    \For{Each service profile $p$ on node $d$}
    \State $j\leftarrow (i, d, r, p)$, $\tau_j=\tau_p, q_j=q_p$, $ c^{\text{GPU}}_{j}=c^{\text{GPU}}_{p}$, 
    \State $c^{\text{CPU}}_{j}=c^{\text{CPU}}_{p}$, $c^{\text{RAM}}_{j}=c^{\text{RAM}}_{p}$, $c^{\text{BW}}_{j}=c^{\text{BW}}_{p}$, 
    \If{ $ |r_j|\leq h_{p} \And |r_j|\leq \hat{h}_{i} \And \tau_{p}\leq\hat{\tau}_i\And q_{p}\geq \hat{q}_i \And c^{\text{BW}}_{j}\leq \min_{e\in {r}}c^{\text{BW}}_{e} $}
        \State $\ccalJ_i\leftarrow\ccalJ_i\cup\{j\}$
    % \State Discard service profiles with bandwidth consumption exceeding available bandwidth along any extracted route.
    % \State Associate $d$'s filtered service profiles with valid routes starting from $s$, referring to them as offloading options.
    \EndIf
    \EndFor
    \EndFor
\EndFor

\State \Return $\ccalJ_i$.
\end{algorithmic}
\end{algorithm}

The objective of \eqref{eq:mip} is to maximize the total utility of assignment $\bbx$, as defined in \eqref{eq:mip:obj}, where $u_j$ is the utility of offloading option $j$ as the weighted sum of latency and accuracy, and $0\leq \lambda \leq 1$ is a tunable parameter that controls the trade-off between latency and accuracy.
By adjusting $\lambda$, users can prioritize computational speed or result quality based on specific application requirements.
The MIP formulation in~\eqref{eq:mip} is NP-hard due to the integer constraints in~\eqref{eq:mip:binary}, for which finding an exact solution requires a computational complexity exponential to the scale of the problem.

% \textcolor{red}{SS: `For example, $c^{\text{CPU}}_{j,v}$ encodes ... while  $C^{\text{CPU}}_v$ represents .... Considerations apply to RAM and bandwidth (BW).' (something like that) Also, we need to explain why we have (3) there ... here is where the wired vs wireless thing might come to play. Something like "In terms of bandwidth, the constraint in (2) does not consider the interaction of different flows going through the same edge. In many cases, especially in wireless links, interference effects might lead to underutilization of the edge bandwidth. To account for this, in (3) we set an constraint on the global bandwidth ..., something like that.}
% The task assignment constraint in \eqref{eq:task_assignment} guarantees that each task is assigned to exactly one resource workflow. 
% Furthermore, the binary nature of the decision variables in \eqref{eq:mip:binary} reflects the discrete nature of the task allocation process.
% \textcolor{red}{SS: you have already defined the acronym MIP, here you can just use it} 
% \textcolor{red}{SS: The solvers are not NP-hard, the problem is NP-hard.} 
% To mitigate this computational burden, we introduce our Sparse-Lagrangian Continuous Relaxation technique in the following section.

\section{Iterative Convex Relaxation}\label{sec:method}
To mitigate the NP-hard complexity associated with integer decision variables in our optimization problem, we first relax the binary constraints in \eqref{eq:mip:binary}, allowing the decision variables to take continuous values while ensuring they sum to one across each task:
\begin{align}\label{eq:cont_variables}  
   0\leq x_{j} \leq 1, \; \forall \; j \in \ccalJ.  
\end{align}  
This relaxation transforms the problem into a convex form, significantly reducing the computational burden of MIP 
% \textcolor{red}{SS: Again, just use the acronym that you have already defined}
solvers. 
However, it compromises the strict enforcement of the constraint in \eqref{eq:mip:binary}.
To guide the relaxed variables toward integer values while maintaining computational efficiency, we adopt the reweighted $L_1$-minimization 
% \textcolor{red}{SS: You call this sometimes reweighted $L_1$-normalization and sometimes reweighted $L_1$-minimization while you refer to the same thing. This is confusing. Simply call it minimization if you refer to the same thing} 
technique in \cite{candes2008enhancing}, which promotes sparsity in the continuous decision variables. 
To integrate this approach into our proposed algorithm, we also employ the
% \textcolor{red}{SS: What do you mean by `to ensure the effectiveness'?} 
primal-dual optimization in \cite{kochenderfer2019algorithms} to relax the computational constraints in \eqref{eq:mip:cpu}, \eqref{eq:mip:ram}, \eqref{eq:mip:bw} and \eqref{eq:mip:gbw}. 
Instead of enforcing these constraints as \textit{hard} constraints to the solver, we incorporate them into the total utility in \eqref{eq:mip:obj} through Lagrange dual variables.
The details of these two key techniques are discussed in the following.
% \begin{align}
%     &\min_{x_{j}} \; \sum_{i\in\ccalT} \sum_{j \in \mathcal{J}_i} -[ (1-\lambda )\cdot \text{accuracy}(j) - \lambda \cdot \text{latency}(j) ] x_{j}  \nonumber \\ \nonumber
%     &\text{subject to} \\ \nonumber
%     &\sum_{i\in\ccalT} \sum_{j \in \mathcal{J}_i} c^{\text{CPU}}_{j,v}x_{j}-C^{\text{CPU}}_v \leq 0 \; , \; \; \; \forall v \in \mathcal{V} \\ \nonumber
%     &\sum_{i\in\ccalT} \sum_{j \in \mathcal{J}_i} c^{\text{RAM}}_{j,v}x_{j}-C^{\text{RAM}}_v \leq 0 \; , \; \; \; \forall v \in \mathcal{V} \\
%     &\sum_{i\in\ccalT} \sum_{j \in \mathcal{J}_i} c^{\text{BW}}_{j,e}x_{j}-C^{\text{BW}}_e \leq 0 \; , \; \; \; \forall e \in \mathcal{E} \\ \nonumber
%     &\sum_{i\in\ccalT} \sum_{j \in \mathcal{J}_i}\sum_{e \in \mathcal{E}} c^{\text{BW}}_{j,e}x_{j}-C^{\text{BW}} \leq 0 \; ,\\ \nonumber
%     &\sum_{j \in \mathcal{J}_i}x_{j}=1 \; \; \; \forall n \in \{1, ..., N\}\\ \nonumber
%     &x_{j} \in \{0,1\}  \; \; \; \forall n \in \{1, ..., N\}\; , \;j \in \mathcal{J}_i
% \end{align}

% \subsection{Convex Optimization Problem in Offloading}

\subsection{Reweighted $L_1$-Minimization}\label{subsec:l1norm}
Reweighted $L_1$-minimization \cite{candes2008enhancing} is a method designed to approximate the $L_0$-norm by iteratively refining the $L_1$-norm through adaptive weighting, making it a powerful tool for promoting sparsity. 
In the context of our task allocation problem, this approach is applied to encourage the relaxed continuous decision variables $x_{j}$ in \eqref{eq:cont_variables} to converge toward their integer counterparts, thereby approximating the original non-convex optimization problem efficiently.

We define utility vector $\bbu=[u_j\mid j\in\ccalJ]$.
The relaxed optimization problem with a (non-weighted) $L_1$ regularizer to promote sparsity is given by:
\begin{align}\label{eq:l1norm}
    \bbx^* &= \argmin_{\bbx}\;\bbx^{\top}\bbu+\gamma \|\mathbf{x}\|_1, 
    \\
    \text{s.t. }\;\;\; &  \enspace \eqref{eq:mip:cpu}, \eqref{eq:mip:ram}, \eqref{eq:mip:bw}, \eqref{eq:mip:gbw}, \eqref{eq:task_assignment}, \text{ and 
 }\eqref{eq:cont_variables} \nonumber
\end{align}
where $ \|\cdot\|_1 $ stands for $L_1$ norm, and $\gamma$ is a scalar constant that controls the trade-off between the original objective and the sparsity-promoting term $\|\mathbf{x}\|_1$. 
% To reduce complexity, we fix $\gamma = 1$ across all experiments.
Instead of \eqref{eq:l1norm}, we employ an iterative refinement of $\|\mathbf{x}\|_1$ using the reweighted $L_1$-minimization (RL1) algorithm.
The RL1 algorithm promotes sparsity in the assignment variables $x_{j}$ by iteratively updating the weights associated with the $L_1$ term. 
The procedure is as follows:

\noindent
\textbf{1. Initialization:} Initialize the weights $w_{j}{(1)} = 0$ for all $j \in \ccalJ$, and set an iteration limit $K$ and stopping criterion.

\noindent
\textbf{2. Iterative Update:} For each iteration $k = 1, 2, \dots, K$, solve the following convex optimization: 
% \textcolor{red}{SS: Here notice that in the objective you should have $x_{i,j}$ instead of $x^{(k)}_{i,j}$, right? that should be the optimization variable not the solution from the previous iteration. The solution of the previous iteration only appears through the weights.}
\begin{align}
   \bbx{(k+1)} &=  \argmin_{\bbx}\; \bbx^{\top}\bbu + {\gamma} \cdot \bbx^{\top}\bbw(k)  \\
    \text{ s.t } \;\;\; & \nonumber  \eqref{eq:mip:cpu}, \eqref{eq:mip:ram}, \eqref{eq:mip:bw}, \eqref{eq:mip:gbw}, \eqref{eq:task_assignment}, \text{ and 
 }\eqref{eq:cont_variables},  
\end{align}
and update weight vector $\bbw(k+1)=[w_j(k+1)\mid j\in\ccalJ]$, where
\begin{equation}\label{eq:weights}
 w_{j}{(k+1)} =  \frac{1}{|x_{j}{(k+1)}| + \epsilon} ,\; 1\gg \epsilon>0,    
\end{equation}
and $\epsilon$ is a small constant for numerical stability. This step increases the penalty for smaller coefficients, iteratively driving non-zero values of $x_{j}$ toward sparsity.

\noindent
\textbf{3. Termination:} The algorithm stops after $K$ iterations or when the change in $\mathbf{x}{(k)}$ between successive iterations is negligible.
% \textcolor{red}{SS: Notice that I changed the following in several places as in the previous expression. To bold a mathematical object you should use mathbf instead of bf in latex} 

By incorporating RL1, the relaxed task allocation problem retains the computational efficiency of a convex optimization framework while steering the solution toward sparsity, ensuring that tasks are effectively assigned to a single offloading option. The sparse solutions achieved through RL1 closely approximate the binary constraints in \eqref{eq:mip:binary} without resorting to computationally expensive MIP solvers.

\subsection{Primal-dual Optimization}\label{subsec:lagrange}

When addressing the task allocation problem, the resource constraints in \eqref{eq:mip:cpu}, \eqref{eq:mip:ram}, \eqref{eq:mip:bw} and \eqref{eq:mip:gbw} can be incorporated directly into the total utility in \eqref{eq:mip:obj}.
% without relying on solver-based hard constraints.
% \textcolor{red}{SS: The meaning of this previous sentence is unclear. What do you mean to incorporate the constraints in the optimization framework?}. 
This is achieved through the primal-dual optimization method, which transforms the original constrained optimization problem into an unconstrained one by introducing Lagrange multipliers.
% For simplicity in this section we represent the decision variables in vector form.
Given our constrained optimization problem in \eqref{eq:mip}, the Lagrangian is  defined as: 
\begin{align}\label{eq:Lagrange}
    \mathcal{L}(\mathbf{x}, \mu) =  
    &\bbx^{\top}\bbu + \sum_{v\in\mathcal{V}}\mu_{v}^{\text{CPU}}\left[\sum_{j \in \mathcal{J}} c^{\text{CPU}}_{j,v}x_{j} - C^{\text{CPU}}_v \right] \nonumber\\ 
    & + \sum_{{v\in\mathcal{V}}}\mu_{v}^{\text{RAM}}\left[
   \sum_{j \in \mathcal{J}} c^{\text{RAM}}_{j,v}x_{j} - C^{\text{RAM}}_v\right]\\ \nonumber
   & + \sum_{e\in \mathcal{E}}\mu_{e}^{\text{BW}}\left[\sum_{j \in \mathcal{J}} c^{\text{BW}}_{j,e}x_{j} - C^{\text{BW}}_e\right]\\ \nonumber  
   & +\mu^{\text{GBW}}\left[\sum_{j \in \mathcal{J}} \sum_{e \in \mathcal{E}} c^{\text{BW}}_{j,e}x_{j} - C^{\text{GBW}}\right], 
\end{align}
% \[
% \mathcal{L}(x, \lambda) = f(x) + \sum_{l=1}^{M} \lambda_l g_l(x),
% \] 
where vector $\mu = \left[[\mu_{v}^{\text{CPU}}, \mu_{v}^{\text{RAM}}]_{v\in\ccalV}, [\mu_{e}^{\text{BW}}]_{e\in\ccalE}, \mu^{\text{GBW}}\right] \geq 0$ collects the Lagrangian multipliers that penalize violations of all the constraints.
By minimizing \eqref{eq:Lagrange} with respect to the primal variables $\mathbf{x}$ the method yields the dual function: 
\[
g(\mu) = \inf_{\mathbf{x}} \mathcal{L}(\mathbf{x}, \mu),
\] 
and the corresponding dual problem is: 
\begin{equation}\label{eq:dual}
\mu^*=\argmax_{\mu \geq 0}\; g(\mu).    
\end{equation}

To solve \eqref{eq:dual}, we use the dual ascent method which is an iterative approach.
Starting with an initial guess (typically 0) for the multipliers $\mu$, the dual variables are updated at each iteration as: 
\begin{align}\label{eq:dascent}
    \mu{(k+1)} = \max \left[0, \mu{(k)} + \alpha \nabla_{\mu(k)} g(\mu{(k)})\right],
\end{align}
where $\nabla_{\mu(k)} g(\mu{(k)})$ is the partial derivative of the dual function with respect to $\mu{(k)}$ which accounts for the constraint violation at iteration $k$, and $\alpha$ is a step size. 
This iterative process continues until convergence, ensuring that the dual variables align with the constraints of the original problem.

\section{Sparsity-enhanced Lagrangian Relaxation}\label{sec:sparselagrange}
Our proposed SeLR approach integrates the iterative methods from Sections \ref{subsec:l1norm} and \ref{subsec:lagrange}, which iteratively solves the following convex optimization:
% \textcolor{red}{SS: The definition here of the variables is incorrect since the weight in (8) you defined it as depending on the solution at iteration $k$ and here you are using those weights to find the solution at iteration $k$. I would make it more similar to (8). I will put it as I think it should be, I did not correct the objective function in (8), though. I'll leave it for you.}
\begin{subequations}\label{eq:SparseLagrange}
\begin{align}
    \bbx{(k+1)} =& \argmin_{\bbx}\; \ccalL(\bbx,\mu(k)) + {\gamma} \cdot \bbx^{\top}\bbw(k)\\
    \text{s.t. }\;\;\; & \eqref{eq:task_assignment}, \eqref{eq:cont_variables}, \eqref{eq:weights}, \eqref{eq:dascent}\;,\nonumber\\
    \bbx(k)-\delta&\leq \bbx(k+1)\leq \bbx(k)+\delta\;, \label{eq:SL:stepsize}
\end{align}    
\end{subequations}
where \eqref{eq:SL:stepsize} limit the step size of each iteration to ensure stability.
This convex relaxation convert NP-hard MIP problem into linear programming, employing sparsity regularization to gradually drive decision variables into an integer solution. 
Through simultaneous dual ascent and reweighting, our approach expands the search space of RL1 algorithm from rigid, irregular constrained space to larger continuous space, allowing reaching a near-global optimum through a smoother path. 
Additionally, the dual ascent method in \eqref{eq:dascent} ensures resource constraints are maintained. 
This iterative approach balances flexibility, feasibility, and optimization, achieving near-optimal solutions with lower computational complexity than direct MIP solving, despite requiring iterative updates.

To accelerate the offloading assignment, we further introduce Algorithm \ref{alg:task_offloading} to iteratively assign tasks based on partial integer solutions and reduce the size of the remaining problem, until all tasks are assigned with an offloading option.
In lines 7-8, we solve a convex relaxation of~\eqref{eq:mip} by replacing binary constraint \eqref{eq:mip:binary} with continuous box constrain~\eqref{eq:cont_variables}, bootstrapping $\bbw(k+1)$ for SeLR in line~10 with a valid solution.
% We present the pseudo-code for our proposed iterative computation offloading approach in Algorithm \ref{alg:task_offloading}.

\begin{algorithm}[!t]
\caption{Task Offloading via Iterative Convex Relaxation}
\label{alg:task_offloading}
\begin{algorithmic}[1]

\Require $\ccalG=(\ccalV,\ccalE), \ccalT, K, \delta$
\Ensure Output task offloading assignment $\hat{\bbx}$

\State Initialize $\{\ccalJ_i\}_{i\in\ccalT}$ using Algorithm~\ref{alg:rworkflow_construction}
\State $\mu(1)\gets 0$, $\bbw(1)\gets 0 $, $k\gets 1$, $\hat{\bbx}=\{0\}^{|\ccalJ|}$, $\xi\gets \emptyset$
% \State 

\While {$\ccalT\neq \emptyset$ and $k\leq K$}
    % \State $Allocated \gets \text{False}$
    \State $\ccalJ\leftarrow\bigcup_{i\in\ccalT}\ccalJ_i $; $\bbu\gets[u_j\mid j\in\ccalJ] $
    \State Remove $[w_j(k)|j\in\xi]$ from $\bbw(k)$ so that $\bbw(k)\in\reals^{|\ccalJ|}$
    % \State Add variables to solver using \eqref{eq:task_assignment} and \eqref{eq:cont_variables}
    \If {$k = 1$ or $\xi\neq\emptyset$}
        \State Get $\bbx(k+1)$ by solving a convex relaxation of \eqref{eq:mip} that replaces~\eqref{eq:mip:binary} with~\eqref{eq:cont_variables}
        \State Reset Lagrange multipliers $\mu(k) \gets 0$
    \Else
        \State Get $\bbx(k+1)$ by solving \eqref{eq:SparseLagrange} with $\bbw(k)$
        % \State To avoid sudden jumps set $|x_{j}{(k)}-x_{j}{(k-1)}| \leq 0.1$
    \EndIf

    \State $\xi\gets \emptyset$

    \For {$i\in\ccalT$}
    \If{$\exists\; x_j(k+1)=1, j\in\ccalJ_i $}
        \State Assign task $i$, $\hat{x}_j\gets x_j(k+1)$
        \State Update $\ccalJ$ by removing offloading options no longer supported by the residual capacities after assigning task $i$.
        \State $\ccalT\gets \ccalT\setminus\{i\}$,\; $\xi\gets \xi\cup\{j\}$
    \EndIf
    \EndFor
    \If{$\xi=\emptyset$}
        \State Compute $\bbw(k+1)$ using \eqref{eq:weights}
       % \State Store solution values $\bbx{(k+1)}$ to guide the next iterations
        \State Update $\mu(k+1)$ via dual ascent in \eqref{eq:dascent}
        % \State Check for resource violations
        % \If {violations exist}
        % \Else
        %     \State \textbf{Break} loop if no tasks remain unassigned
        % \EndIf
    \EndIf
    \State $k \gets k + 1$
\EndWhile

% \State \textbf{Return} performance metrics

\end{algorithmic}
\end{algorithm}

\section{Numerical Experiments}\label{sec:results}
We evaluate our proposed approach and other benchmarks numerically with simulated tasks and multi-hop networks. 
% \textcolor{red}{SS: No need to say that there was another dataset that we cannot release. You can directly say something like: We generated all datasets --- including tasks, workflows, network topology, and resource attributes --- inspired by battlefield edge networks and supported by network simulators. or something like that}
All test instances --- including tasks, service profiles, network topology, and resource attributes --- are inspired by tactical edge networks and supported by network simulators.
% The evaluations in this section include comparisons with several benchmark task allocation schemes.
\footnote{Source code at \url{https://github.com/Negar-Erfanian/SeLR}.
}
% \textcolor{red}{SS: Why are we talking in future tense? I guess this is because we need green light from the army to make it publicly available, right?
% You can still say that the code is available and put a link to a repo in a footnote, even if that repo is empty right now, and we can get the authorization from them and then make it public.}

% \begin{figure*}[t!]
%     % \hspace{-3mm}
%     \subfloat[]{
%     % \includegraphics[width=0.95\linewidth]{Figures/Mixed_ErrorbarDelayResults_randomprob0.5.pdf}
%     \includegraphics[width=0.33\linewidth]{samples/histogram_40tasks.pdf}
%     \label{fig:hist40}
%     }
%     % \hspace{-3mm}
%     \subfloat[]{

%     \includegraphics[width=0.33\linewidth]{samples/histogram_70tasks.pdf}
%     \label{fig:hist70}
%     }
%     % \hspace{-3mm}
%     \subfloat[]{
%     \includegraphics[width=0.33\linewidth]{samples/histogram_90tasks.pdf}\label{fig:robust}\label{fig:hist90}
%     }  
%     \vspace{-0.1in}
%     \caption{} 
%     \label{fig:hist}
%     \vspace{-0.1in}
% \end{figure*}

\subsection{Test Setup}\label{sec:results:config}

\begin{table}
  \caption{Test Configurations}
  \label{tab:config}
  \begin{tabular}{p{1.3in} | p{1.7in}}
    \toprule
    Configuration & Value or distribution \\
    \midrule
    Network size $|\ccalV|$ & 300 \\
    Sensor CPU capacity &  $2000 $ MIPS \\ 
    Sensor RAM capacity & Uniform random in $\{2.8, 2.9\}$ GB \\ 
    Server CPU capacity & $6000 $ MIPS \\ 
    Server RAM capacity & $25\%$ core servers: $13+\mathbb{U}(0.1,1)$ GB, the rest: $[15,15.3]$ GB  \\
    Link capacity & $ C^{\text{BW}}_e=7 \text{ Mbps}, e\in\ccalE $  \\
    Chance of GPU availability & $100\%$ for core servers, $50\%$ for edge servers, $0\%$ for sensors  \\
    Global bandwidth limit & $ C^{\text{GBW}}=9 $ Gbps  \\
    Max-hop & $ \hat{h}_i=h_{p}=2, \forall\; i\in\ccalT $, for all $p$ \\
    Required task accuracy & $\hat{q}_i\in\mathbb{N}(60,0.1)$ \\
    Required task latency & $ \hat{\tau}_i\in\mathbb{N}(1,0.1)$ \\
    $D$ \# samples per task & $5000$ \\
    $\epsilon$ & 0.0001 \\
    $\delta$ & 0.1\\
    $\alpha$ & 1.0\\
    $\gamma$ & 1.0\\
    $\lambda$ & $0.5$ unless otherwise stated  \\
  \bottomrule
\end{tabular}
\end{table}

The configurations of our numerical experiments are detailed in Table~\ref{tab:config}.
Random networks with $|\mathcal{V}| = 300$ nodes and hierarchical structures, similar to the experiments in~\cite{zhao2025icassp} and exemplified in Fig.~\ref{fig:main}, are used in simulations. 
The network topology is formed in 3 steps:
First, we create $5\%\sim10\%|\ccalV|$ core servers with the highest computational capacities and 100\% GPU availability, forming a clique.
Then, we add $20\%\sim30\%|\ccalV|$ edge servers grouped around each core server, where edge servers within the same group and the corresponding core server are fully connected to each other, whereas edge servers in different groups have no direct connections. 
The capacities of edge servers are less abundant, and the GPU availabilities are set at a probability of $0.5$.
Lastly, we add the rest of nodes as sensors, where each sensor is connected to an edge server, and with a probability of $0.1$ to a core server of the same group. 
Sensors are not directed connected to each other. 
The units of RAM and CPU are respectively gigabytes (GB) and millions of instructions per second (MIPS).
A total of 30 random network instances are used.

% \textcolor{red}{SS: These proportions seem to be too small. Please double check. These are probably proportions and not percentages? or at least they are 2\% instead of 0.2\%?}. 
\begin{table*}[!t]
    \centering
    \caption{Simulated Service Profiles, \{XXX, YYY\} indicates uniformly random selection of one of two values }
    \label{tab:simulated_workflows}
    \resizebox{\textwidth}{!}{%
    \begin{tabular}{ccccccccccccc}
        \toprule
        Model & Accuracy (\%) & GPU & Max-Hops & CPU usage (MIPS) & Exec Time (s) & Load Time-CPU (s) & Load Time-GPU (s) & RAM Util-CPU (MB)& RAM Util-GPU (MB) & Bandwidth (Mbps) & Device Type \\
        \midrule
        M\_1  & $\mathbb{U}(66, 72)$  & \{True, False\} & 2 & 4500 & $\mathbb{U}(0.039, 0.161)$ & $\mathbb{U}(3.63, 5.02)$ &$\mathbb{U}(0.031, 0.048)$  & $\mathbb{U}(2853, 2900)$ & $\mathbb{U}(206, 309)$  & 5  & Server  \\
        M\_2 & $\mathbb{U}(63, 69)$  & \{True, False\} & 2 & 4000 & $\mathbb{U}(0.042, 0.134)$ & $\mathbb{U}(3.52, 4.14)$ & $\mathbb{U}(0.03, 0.073)$& $\mathbb{U}(2854, 2922)$ & $\mathbb{U}(193, 347)$ & 5  & Server  \\
        M\_3 & $\mathbb{U}(62, 69)$  & \{True, False\} & 2 & 3500 & $\mathbb{U}(0.047, 0.157)$ & $\mathbb{U}(3.34, 3.97)$ &  $\mathbb{U}(0.028, 0.077)$ & $\mathbb{U}(2855, 2867)$ & $\mathbb{U}(190, 290)$ & 5  & Server  \\
        M\_4 & $\mathbb{U}(57, 63)$  & \{True, False\} & 2 & 3000 & $\mathbb{U}(0.048, 0.161)$ & $\mathbb{U}(3.35, 4.23)$ & $\mathbb{U}(0.025, 0.062)$ & $\mathbb{U}(2854, 2910)$ & $\mathbb{U}(180, 212)$ & 5  & Server  \\
        M\_5 & $\mathbb{U}(50, 56)$  & \{True, False\} & 2 & \{2500,4500\} & $\mathbb{U}(0.048, 0.155)$ & $\mathbb{U}(3.66, 3.98)$ & $\mathbb{U}(0.023, 0.064)$ & $\mathbb{U}(2851, 2869)$ & $\mathbb{U}(177, 194)$ & 5  & Server  \\
        M\_6 & $\mathbb{U}(47, 52)$  & \{True, False\} & 2 & \{2000,4000\} & $\mathbb{U}(0.057, 0.147)$ & $\mathbb{U}(3.41, 4.16)$ & $\mathbb{U}(0.024, 0.071)$ & $\mathbb{U}(2854, 2869)$ & $\mathbb{U}(180, 202)$ & 5  & Server  \\
        M\_7 & $\mathbb{U}(45, 51)$  & \{True, False\} & 2 & \{4500,3500\} & $\mathbb{U}(0.039, 0.159)$ & $\mathbb{U}(3.68, 4.06)$ & $\mathbb{U}(0.022, 0.068)$ & $\mathbb{U}(2857, 2876)$ & $\mathbb{U}(179, 190)$ & 5  & Server  \\
        \midrule
        M\_8  & 72  & False & 0 & 1000 & 0.123 & 0 & 0.15 & 0 & 461  & 0  & Sensor  \\
        M\_9 & 70  & False & 0 & 1000 & 0.138 & 0 & 0.16 & 0 & 464  & 0  & Sensor  \\
        M\_10 & 69  & False & 0 & 1000 & 0.105 & 0 & 0.14 & 0 & 448  & 0  & Sensor  \\
        M\_11 & 63  & False & 0 & 1000 & 0.087 & 0 & 0.10 & 0 & 441  & 0  & Sensor  \\
        M\_12 & 56  & False & 0 & 1000 & 0.064 & 0 & 0.11 & 0 & 427  & 0  & Sensor  \\
        M\_13 & 52  & False & 0 & 1000 & 0.076 & 0 & 0.12 & 0 &  434  & 0  & Sensor  \\
        M\_14 & 51  & False & 0 & 1000 & 0.064 & 0 & 0.10 & 0 & 439  & 0  & Sensor  \\
        \bottomrule
    \end{tabular}%
    }
\end{table*}

To evaluate the scalability of tested schedulers across different workloads, on each network instance, we create $100$ test instances by generating $1\leq|\ccalT|\leq 100$ tasks on a fixed set of $100$ randomly selected sensor nodes.
Tasks are added incrementally, so that the test instance with $|\ccalT|=N+1$ tasks always include the same set of tasks in the test instance with $|\ccalT|=N$.
This incremental setup ensures a fair and consistent performance comparison across workloads. 
Each task has randomly sampled minimal required accuracy and maximal allowable latency in Table~\ref{tab:config}.
The service profiles are detailed in Table~\ref{tab:simulated_workflows}.
For utility function $u_j$ in \eqref{eq:mip:obj}, we set $\lambda=0.5$ unless otherwise stated,
with a per-sample latency profile $\tau_p$ for utility function is estimated as Exec Time + Load Time/$D$, where $D=5000$ represents number of data samples per task, and the accuracy profile follows uniform distributions as listed in Table~\ref{tab:simulated_workflows}.
% Required latency per task is sampled from a Gaussian distribution with a mean of 1, standard deviation of 0.1.
% , and values restricted to $(0, \infty)$ \textcolor{red}{SS: 0 is 10 standard deviations away from the mean, probably no need to truncate in practice, right? The chances of sampling something 10 standard deviations away from the mwan is of the order of $10^-23$.}. 
% The required accuracy per task is sampled from a Gaussian distribution with a mean of 60, standard deviation of 0.1.
% , and values restricted to $(55, 75)$ \textcolor{red}{SS: Again, and here even more, the limits are 50 or more standard deviations away from the mean. You would never sample something outside of those ranges. I guess here someone might raise the question why is the standard deviation so small. 99\% of tasks would have accuracies between 59.7 and 60.3 (i.e., plus minus 3 standard deviations)}. 
Accuracy and latency values are scaled to the same numerical range so that they are weighted similarly with $\lambda=0.5$.

\begin{figure}[!t]
    \includegraphics[width=0.98\linewidth]{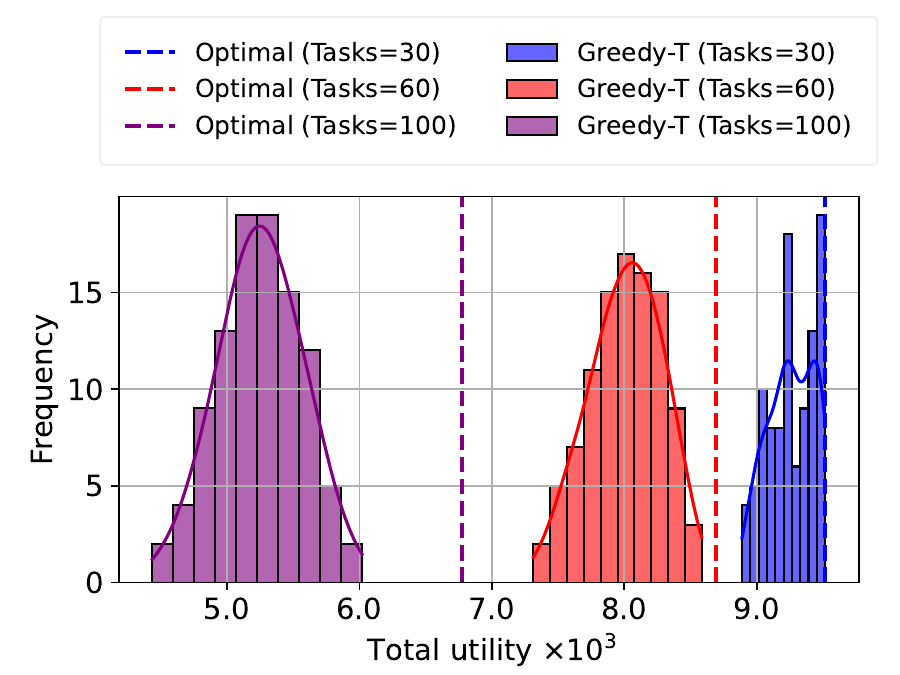}
    \vspace{-0.1in}
    \caption{Utility value distribution for Greedy-T with 100 random order permutations compared to the Optimal utility value for task loads of 30, 60, and 100. 
    The Greedy-T$\times100$ chooses the closest outcome to the Optimal utility.
    % While OrderedGreedy-100 selects the closest outcome to the optimal, we observe that as the task load increases, the distribution, resembling a Gaussian, diverges further from the optimal. 
    % This trend highlights the computational challenges associated with optimized resource allocation using the OrderedGreedy approach in large-scale, complex edge networks with many users. 
    The utility function is based on $\lambda = 0.5$. 
    % Despite the small difference between Greedy and Optimal in terms of objective value, which is due to the scaling of accuracy and latency, the main takeaway is to demonstrate the computational expense and inefficiency of performing Greedy $M$ times, particularly under high task loads.
    % \textcolor{purple}{ZZ: Please adjust the shape and font size of the figure.}
    }
    \label{fig:hist}
    \vspace{-0.1in}
\end{figure}

For comparison, we also evaluate the following approaches.
1)~\textit{\textbf{Optimal}}: Having integer decision variables, we use MIP
% \textcolor{red}{SS: Use the acronym}
to solve the main non-convex optimization problem using the SCIP solver from OR-Tools~\cite{ortools}.
This approach provides the globally optimal solution to the task allocation problem.
2)~\textit{\textbf{Linear-Relax}}: 
We find a fractional solution by relaxing decision variables into continuous values between 0 and 1. 
Tasks are then assigned deterministically to feasible offloading options based on the highest solution values, with resource availability dynamically updated after each allocation until all tasks are placed.
For this approach, we utilized the GLOP solver from OR-Tools~\cite{ortools}, which is specifically designed for convex optimization problems.
3)~\textit{\textbf{SeLR (our approach)}}: Our method relaxes the decision variables to continuous values between 0 and 1, and incorporates iterative sparsification and dual ascent processes to find integer solutions, as specified in Algorithm~\ref{alg:task_offloading}.
We used the same solver as in Linear-Relax.
4)~\textit{\textbf{Greedy-U}}: This method check the feasibility of each offloading option in $\ccalJ$ in the descending order of their utility values.
If an offloading option is feasible based on residual capacities in the server and network, then we assign it to the corresponding task, disable other offloading options of that task.
Otherwise, if an offloading option is infeasible due to violation of constraints on residual capacities, it is invalidated and will not be visited again.
Once a task is assigned an offloading option, the residual capacities in the corresponding servers and links are updated.
This process terminates when all tasks are allocated.
% , operates without the use of a built-in solver, instead relying on the utility value of each offloading option for every task.
% \textcolor{red}{SS: This previous sentence is confusing, what do you mean that you make all variables equal to 1?}.
% Tasks are allocated deterministically based on the offloading option with the highest utility among all tasks, without enforcing a specific allocation order.
% \textcolor{red}{SS: This previous sentence is also unclear}.
% Once an offloading option with the highest utility is selected, its feasibility is assessed. 
% If the allocation is infeasible, it is ignored in subsequent iterations; otherwise, the task assigned to that offloading option will be allocated, and resources are updated accordingly. 
5)~\textit{\textbf{Greedy-T}}: This method assigns tasks sequentially based on a given random permutation of tasks.
For each task, it checks the feasibility of each offloading option based on their utility values in descending order.
The first feasible option is assigned to the task, and then residual capacities are updated accordingly. 
% If not, the task selects the next highest utility option and continues this process until a feasible allocation is found. 
The method will only move on to the next task when the previous task is assigned, therefore is heavily influenced by the given order of tasks. 
% This approach prioritizes feasibility over global optimization and is influenced by the initial task order.
6)~\textit{\textbf{Greedy-U$\times100$}}: Extending the Greedy-T approach, this method evaluates $M$ different permutations of task orderings.
For each permutation, tasks are allocated using the Greedy-T strategy.
The final allocation corresponds to the permutation that achieves the highest overall utility among all $M$ permutations.
Fig.~\ref{fig:hist} illustrates the distribution of total utilities of 100 random permutations of Greedy-T under three different task loads, highlighting how likely Greedy-T$\times$100 can approximate the optimal solution at different problem scales. 
Notice that the Greedy-T$\times$100 also increases computation time by 100 times over Greedy-T. 

% \begin{figure*}[t!]
%     \hspace{-3mm}
%     \subfloat[]{
%     % \includegraphics[width=0.95\linewidth]{Figures/Mixed_ErrorbarDelayResults_randomprob0.5.pdf}
%     \includegraphics[width=0.33\linewidth]{samples/shiftedobjval_vs_tasknum_mem1_cpu1_bw1_lamb0.5_ortools.pdf}
%     \label{fig:shiftedobjval}
%     }
%     \hspace{-3mm}
%     \subfloat[]{

%     \includegraphics[width=0.33\linewidth]{samples/accuracy_vs_latency_all_lambdas.pdf}
%     \label{fig:accuracy_vs_latency}
%     }
%     \hspace{-3mm}
%     \subfloat[]{
%     \includegraphics[width=0.33\linewidth]{samples/alllocaloffloads.pdf}\label{fig:robust}\label{fig:localoffloads}
%     }  
%     \vspace{-0.1in}
%     \caption{} 
%     \label{fig:mixed}
%     \vspace{-0.1in}
% \end{figure*}

\subsection{Performance Under Varying Number of Tasks and Resource Availabilities}\label{subsec:performancevsload}
\begin{figure}[!t]
    \centering
    \includegraphics[width=1\linewidth]{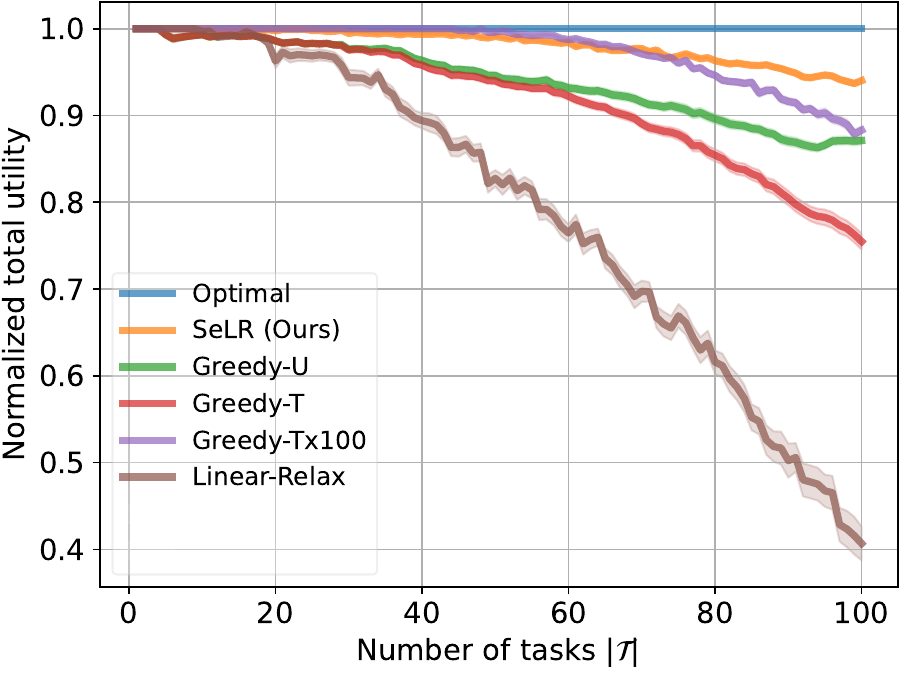}
    \caption{
    % \textcolor{red}{SS: Somehow mark in the legend of the figure which one is our method.} 
    Normalized total utility achieved by the evaluated schedulers as a function of the number of tasks $|\ccalT|$, relative to the optimal solution, with $\lambda = 0.5$. 
    Each curve represents an average over 30 network instances with $|\mathcal{V}| = 300$. 
    On each network instance, tasks are added incrementally  such that the test instance for $|\ccalT| = N+1$ always includes the same set of tasks in the instance with $|\ccalT| = N$ .
    % These results show how well each scheduler performs relative to the optimal solution. 
    % Compared to other heuristics, our approach performs and scales better.
    }
    \label{fig:shiftedobjval}
\end{figure}

To evaluate the scalability of tested schedulers, in Fig.~\ref{fig:shiftedobjval}, we present the total utility achieved by all schemes, normalized by that of the optimal solutions, as a function of the number of tasks.
With a small number of tasks, e.g., $|\ccalT|\leq 15$, all approaches achieve optimal or near-optimal normalized utility.
However, as the number of tasks increases to 60, the normalized total utilities of SeLR, Greedy-T$\times100$, Greedy-U, Greedy-T, and Linear-Relax drop to $ 0.984 $, $0.987$, $0.93$, $0.92$, and $0.77$, respectively.
When the number of tasks increases to $|\ccalV|=100$, these values further decrease to $0.94$, $0.88$, $0.87$, $0.75$, and $0.41$. 
As the problem scales up, the optimality gaps of all tested heuristic schedulers increase, whereas our SeRL suffers the smallest optimality gap at $|\ccalT|=100$.

The significant drop in the performance of the Linear-Relax approach is due to the sparsity of the solution values from the convex solver.
Initially, the method assigns tasks to the highest integer-valued solutions. However, once these are exhausted, it must consider non-integer values. 
Due to resource constraints, most non-integer offloading options become infeasible, forcing the method to choose among zero-valued solutions randomly. 
% This undermines the balance between accuracy and latency, favoring feasibility over optimization. 
This leads to both poor quality of solution as shown in Fig.~\ref{fig:shiftedobjval} and high computational overhead as detailed in Table~\ref{tab:overhead}, since the scheduler must find a feasible solution by iterating over a large number of offloading options with zero solution values.
Due to these limitations, we exclude this approach from the rest of experiments.

Greedy-T and Greedy-T$\times100$ scale similarly; however, Greedy-T$\times$100 improves upon Greedy-T by performing it 100 times and selecting the best outcome.
Nevertheless, it still struggles to scale beyond a certain task load, aligning more closely with the performance of Greedy-U. 
This limitation is due to the combinatorial nature of the task-ordering problem.
Sampling only 100 out of $|\ccalT|!$ possible permutations yields a low probability of approaching the optimal solution, especially as $|\ccalT|$ grows.
This is evident in Fig.~\ref{fig:hist}, where the utility distribution of Greedy-T over 100 permutations gradually diverges from the optimal as the number of tasks increases.
Greedy-U performs similarly to Greedy-T under lower task loads; however, it scales more effectively and eventually approaches the performance of Greedy-T$\times$100 as the problem size increases.

% In contrast, our method consistently remains close to the optimal total utility, maintaining around 95\% of the optimal performance even for 100 tasks.
These results highlight the scalability of our proposed SeLR approach under increasing task loads. 
While all other heuristics face scalability limitations, SeLR exhibits the smallest optimality gap as the problem scales.
Notice that the computational cost of Greedy-T$\times$100 scales linearly with the number of evaluations, as it involves running Greedy-T 100 times. 
Therefore, despite its similar performance to SeLR up to a moderate task load, Greedy-T$\times$100 is significantly more computationally expensive, as reported in Section~\ref{subsec:computationaloverhead}.

% shows the total utility achieved by all schemes, normalized by that of the optimal solutions, as the number of tasks scales up to 100. 
% Performance is presented as a ratio, computed by dividing each scheme’s utility value by that of the optimal solution.

% In this figure, we specifically report the results for $\lambda=0.5$ in~\eqref{eq:mip:obj}, equally weighting accuracy and latency during task placement.
% \textcolor{red}{SS: I don't understand this. Before you say that latency takes mean values of 1 and accuracy mean values of 60. Hence, I would say that dividing by 60 would make more sense. Or dividing by 100 would be ok. But why does dividing by 1000 would give them the same scale? Ah, ok, after looking at Fig. 4 I realized that latencies are actually much smaller than 1 (which is the latency of the task). This might be confusing to other readers too.}

% Greedy-T, followed by Greedy-U, achieve better performance but remain suboptimal. 
% Greedy-T$\times$M, with M = 100, improves upon Greedy-T by evaluating 100 random task orderings and selecting the best outcome. 
% Although it consistently outperforms Greedy-T, and performs comparably to our approach under lighter task loads, it struggles to scale due to the combinatorial nature of the task-ordering problem.
% Sampling only 100 out of $|\ccalT|!$ possible permutations yields a low probability of approaching the optimal solution, especially as $|\ccalT|$ grows.
% This is evident in Fig.~\ref{fig:hist}, where the utility distribution of Greedy-T over 100 permutations gradually diverges from the optimal as the number of tasks increases.
% However, the increasing number of required samples to approach the optimal solution becomes more difficult at higher task loads. 

% \textcolor{red}{SS: I am trying to reconcile Figure 2 and Figure 3. If we take 100 tasks, in Figure 3 it looks like Greedy is around 0.8 of Optimal and Greedy-100 is around 0.9 of optimal. But if liik at the purple histogram and compare a random sample of the histogram (Greedy) and the best sample of the histogram (Greedy-100), I don't see these proporations of optimal (the dashed purple line). So, what am I missing here?}

\begin{figure}[!t]
    \centering
    \includegraphics[width=1\linewidth]{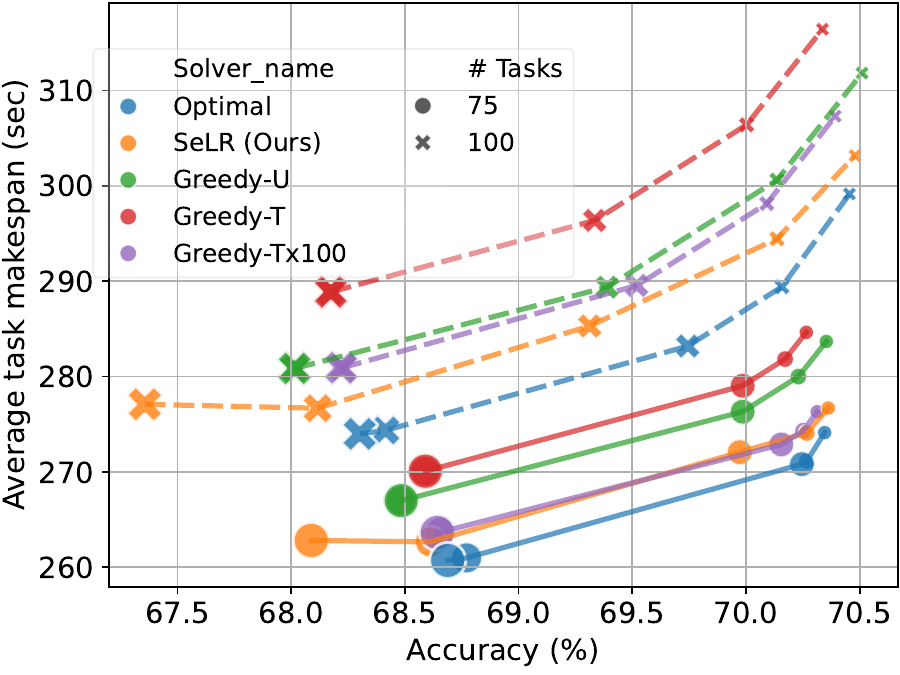}
    \caption{
    % \textcolor{red}{SS: Mark our method in the legend.} 
    Accuracy vs. latency curve for all schemes as a function of gradually increasing $\lambda$ values from 0 to 1. 
    Results are shown for $\lambda \in \{0.1, 0.15, 0.4, 0.6, 0.8\}$, with larger markers indicating higher $\lambda$ values, for task loads of 75 and 100, averaged over 30 network instances with a size of $\mathcal{V} = 300$. 
    For Greedy-U, Greedy-T, and Greedy-T$\times100$, the markers of $\lambda=0.6$ and $\lambda=0.8$ overlap.
    % These results demonstrate how our proposed method scales effectively with problem size, maintaining a close proximity to the optimal curve, compared to other heuristics.
    }
    \label{fig:accuracy_vs_latency}
\end{figure}

To illustrate the trade-offs between accuracy and latency under the tested schedulers, in Fig.~\ref{fig:accuracy_vs_latency}, we present the Pareto frontier of all methods under $\lambda \in \{0.1, 0.15, 0.4, 0.6, 0.8\}$, with accuracy and makespan as $x$ and $y$ axes, in the settings of 75 and 100 tasks. 
Here, $\lambda$ controls the trade-off between accuracy and latency in the utility function, e.g., smaller $\lambda$ prioritizes accuracy over latency, and vise versa.
As shown in Fig.~\ref{fig:accuracy_vs_latency},  larger $\lambda$, as indicated by larger marker, leads to lower latency and accuracy under a given scheduler, while reducing $\lambda$ leads to increased accuracy and latency.
Here, a lower Pareto frontier indicates better performance under all possible trade-offs, where the optimal solver has the lowest frontier. 
For example, if we adjust the $\lambda$ individually for each scheduler so that it achieves an average accuracy of $70\%$, the average latency of these schedulers are the $y$ values with $x=70$ in Fig.~\ref{fig:accuracy_vs_latency}. 
% To avoid overlapping markers, we only display results for $\lambda=\{0.1, 0.15, 0.4, 0.6, 0.8\}$ where the marker size encodes the value of $\lambda$, with larger dots and crosses indicating higher values for the 75-task and 100-task cases, respectively. 
% As $\lambda$ increases, all schemes follow the expected trend along the accuracy-latency trade-off curve, gradually shifting from high-accuracy, high-latency placements towards lower-accuracy, lower-latency ones.

With a workload of 75 tasks, the Pareto frontiers of our SeRL and Greedy-T$\times100$ almost overlap, indicating similar overall quality of solutions despite that SeRL has slightly higher total utility than Greedy-T$\times100$ under $\lambda=0.5$ in Fig.~\ref{fig:shiftedobjval}. 
% However, the trade-off between accuracy and latency results in higher total utility for SeLR compared to Greedy-T$\times$100, as shown earlier in Fig.\ref{fig:shiftedobjval} for $\lambda  = 0.5$. 
As the workload increases to 100 tasks, our SeRL achieves a clearly better Pareto frontier than that of Greedy-T$\times100$, demonstrating better scalability of our approach.
Similarly, Greedy-T also exhibits better scalability than Greedy-U and Greedy-U$\times100$ under various $\lambda$s, which is consistent with the results in Fig.~\ref{fig:shiftedobjval}.

Notice that for smaller values of $\lambda$ (e.g., $\lambda = 0.1$), all approaches exhibit higher accuracy even as the problem scales from 75 to 100 tasks. 
However, as $\lambda$ increases beyond 0.15, this trend shifts, where higher load results in lower accuracy across all.
However, latency always increases as the problem scales for all $\lambda$.
These results indicate that optimizing for latency is more challenging than optimizing for accuracy across all schemes.
For $\lambda = 0.8$ and $0.6$, the accuracy-latency trade-offs overlap across all methods except SeLR at both task loads, and Optimal at the higher task load.
This highlights SeLR’s ability to effectively distinguish the trade-off between accuracy and latency, even with subtle changes in $\lambda$. 
This advantage stems from its iterative design, a limitation in the other schedulers. 
Although similar overlaps occur for many other values of $\lambda$, we excluded them from the figure to maintain visual clarity.
% As depicted, our proposed method closely follows the optimal accuracy-latency curve as the problem scales, whereas other methods diverge from optimal.
% \textcolor{red}{SS: The discussion from here until the end of this paragraph can be deleted. If you want to keep it, we can. But not sure if it adds too much.}
% We conducted the above experiments for $\lambda = \{0, 1\}$, representing the cases of maximizing accuracy alone and minimizing latency alone, respectively. 
% For $\lambda = 1$, the results were very similar to those for $\lambda = 0.8$. However, for $\lambda = 0$, we observed a significant increase in both accuracy and latency, causing all approaches to overlap, appearing as outliers in the figure.  
% For these reasons, we omitted these cases from the figure. 
% This observation suggests that, when latency is not a factor in the problem formulation, all approaches tend to converge on selecting offloading options that maximize accuracy. 
% However, the complexity of the task allocation problem increases across all approaches when latency must be balanced with accuracy.

\begin{figure}[!t]
    \centering
    \includegraphics[width=1\linewidth]{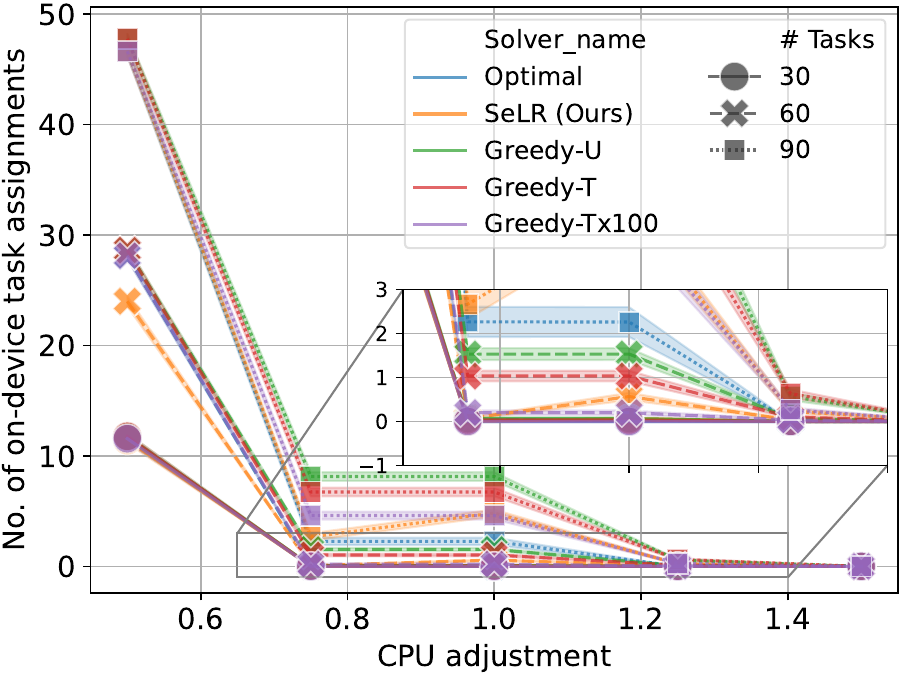}
    \caption{The numbers of on-device task assignments under tested schedulers as the CPU capacities of all nodes scaled by a factor of $\beta\in\{0.5, 0.75, 1, 1.25, 1.5\}$, with other resources such as RAM and bandwidth remaining the same. 
    Workloads $|\ccalT|\in\{30,60,90\}$, and $\lambda = 0.5$. 
    The results are averaged over 30 network instances with $\mathcal{V} = 300$ nodes.
    }
    \label{fig:localoffloads}
\end{figure}

To illustrate how many tasks are offloaded to servers across different workloads, in Fig.~\ref{fig:localoffloads}, we show the number of on-device task assignments as a function of available CPU in resources across all schemes. 
As described in \eqref{eq:mip:cpu}, \eqref{eq:mip:ram}, \eqref{eq:mip:bw}, CPU is one of the key computational resources that must not be exceeded during task placement across servers. 
In these experiments, we vary only the server CPU availability while keeping other resources such as RAM and bandwidth fixed, though the same observations would apply if we varied those as well. 
We evaluate scenarios with 30, 60, and 90 tasks.
To regulate resource availability, we scale the CPU capacities of all servers by a factor of $\beta\in\{0.5, 0.75, 1, 1.25, 1.5\}$. 
These scaling factors are chosen based on our numerical simulations to effectively highlight the sensitivity of the task allocation problem to resource availability.
$\beta$s below 0.5 led to all tasks being executed on-device, while $\beta$s above 1.5 resulted in no on-device assignments; therefore, these cases were omitted from the figure.

As CPU capacity increases, the number of on-device task assignments is generally expected to decrease across all methods, as servers can accommodate more tasks. 
Even though this trend is observed in most cases, there are exceptions, such as with SeLR when $\beta$ increases from 0.75 to 1 for 60 and 90 tasks. 
This deviation can be attributed to the network's complexity, where the optimization process focuses on maximizing overall task allocation utility rather than strictly minimizing on-device executions.
Additionally, we observe that at a fixed CPU capacity, except for $\beta=1.5$, a higher task load consistently results in more on-device executions across all schemes, emphasizing the growing challenge of task offloading as workload increases.
Among all methods, our approach and the optimal solution result in fewer on-device task assignments, especially at larger problem scales, followed by Greedy-T$\times100$, Greedy-U, and Greedy-T in that order.
At $\beta=1$ with 60 tasks, Greedy-T$\times$100 assigns fewer tasks to the device compared to SeLR.
This observation is consistent with the performance comparison shown in Fig.~\ref{fig:shiftedobjval}, where Greedy-T$\times$100 performs slightly better than SeLR under a 60-task load.
Most importantly, the consistent ranking of tested schedulers under varying resource availabilities in this experiment shows that our results in Figs.~\ref{fig:shiftedobjval} and~\ref{fig:accuracy_vs_latency} are representative rather than tied to a specific test configuration.
Like other methods, our SeRL behaves consistently across resource availabilities.

\subsection{Computational Overhead of Schedulers}\label{subsec:computationaloverhead}
\begin{figure}[!t]
    \centering
    \includegraphics[width=1\linewidth]{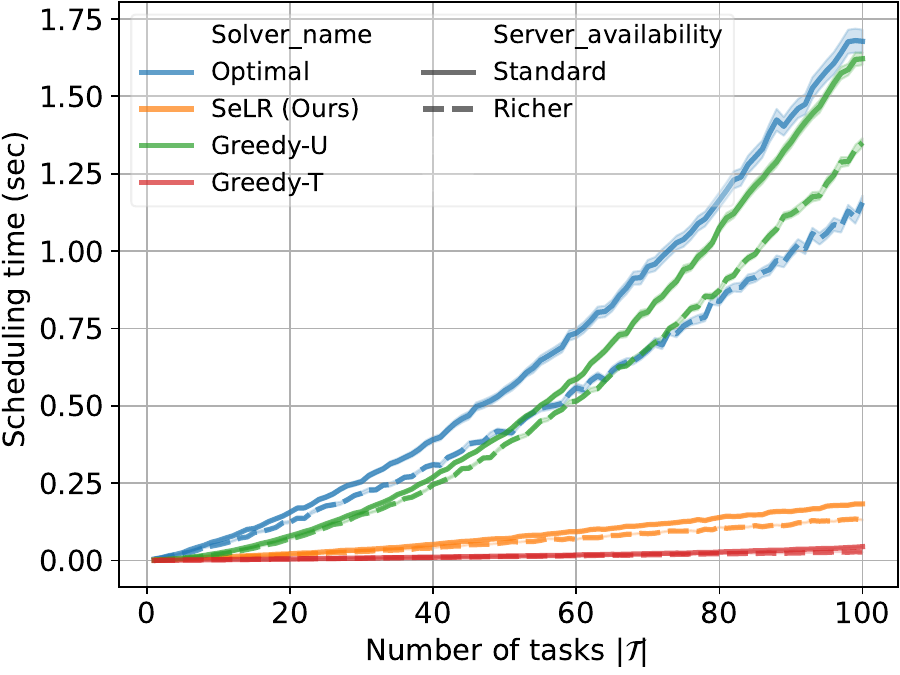}
    \caption{Computing time of four tested schedulers as a function of number of tasks under two settings of server availability in networks of $|\mathcal{V}|=300$ nodes, and $\lambda=0.5$.
    Each point is obtained from the average of 100 instances. 
    The proportions of core servers, edge servers, and sensors in the standard setting are the same as previous experiments described in Section~\ref{sec:results:config}. 
    Under richer server availability, edge servers are increased by $10\%|\ccalV|$ to $30\%\sim40\%|\ccalV|$, and core servers remain the same, making offloading easier with more edge servers.
    % , and the edge server capacity of each core server increases from 6 to 8 when transitioning from the standard to the richer setting.
    % These results show how much faster our proposed method performs compared to the optimal solver as the problem scales in both network settings.
    % Despite its iterative nature, our proposed Cont-relaxed-map method achieves up to 7x faster performance than the Optimal solver, particularly in complex network settings and up to 6x faster in simpler cases with higher task loads.  While Optimal and Rulebased exhibit the highest computational times, Greedy is the most efficient. 
    % However, finding the optimal order to achieve results close to the optimal incurs additional computational cost, making OrderedGreedy-M computationally expensive and therefore, omitted from this figure. \textcolor{red}{SS: The caption here is a bit too long. You already mention that we exclude Greedy-M in the text no need to repeat. You can minimize the redundancy a bit more too.}
    }
    \label{fig:computationaltime}
\end{figure}

To illustrate the computational efficiency of our SeLR, in Fig.~\ref{fig:computationaltime} and Table~\ref{tab:overhead}, we present the computational overhead of tested schedulers as the runtime for finding a solution across different workloads $|\ccalT|$ under two settings of server availabilities in networks of $|\mathcal{V}|=300$ nodes.
% All methods operate on the same offloading options $\ccalJ$. 
% The first two optimize directly over this matrix by defining appropriate decision variables, whereas heuristic methods perform task placement based on the objective values derived from it. 
% This setup ensures a fair evaluation by keeping the task placement framework consistent across all methods.
In the standard server availability setting, the proportions of core and edge servers are configured the same as in previous experiments specified in Section~\ref{sec:results:config}.
In the setting of richer server availability, the proportions of core servers remain the same as the standard setting, while the edge servers is increased to $30\% \sim 40\%$ of $|\ccalV|$, making the task offloading easier with more servers. 
Each point in Fig.~\ref{fig:computationaltime} is the average runtime of 30 instances.

\begin{table}[!t]
  \centering
  \caption{Scheduling overhead (in milliseconds) for 50-100 tasks under standard and richer server availabilities.}
  \label{tab:overhead}
  % \small
  \begin{tabular}{p{2cm}|p{1.1cm}|p{1.15cm}|p{1.1cm}|p{1.15cm}|} % Adjust widths as needed
    \toprule
    \multirow{2}{4em}{\textbf{Scheduler}} & \multicolumn{2}{c|}{Standard} & \multicolumn{2}{c|}{Richer} \\ \cline{2-5}
    & 50 tasks & 100 tasks & 50 tasks & 100 tasks \\
    \midrule
    Optimal            & 548 & 1678     & 415 & 1157   \\
    \textbf{SeLR (Ours)} & \textbf{71} & \textbf{183}    & \textbf{58} & \textbf{132}    \\
    Greedy-T           & 14 & 45        & 13 & 28      \\
    Greedy-T$\times100$ & 1383 & 4418    & 1204 & 2646  \\
    Greedy-U           & 409 & 1621     & 375 & 1350   \\
    Linear-Relax       & 9857 & 287763  & 4421 & 51652 \\
    \bottomrule
  \end{tabular}
\end{table}

The Optimal and Greedy-U schedulers have the highest computational overhead, with 1678 ms and 1621 ms under the standard setting at a 100-task load, and 1157 ms and 1350 ms under the richer setting, respectively.
Greedy-T shows the lowest computational overhead, with computational overhead of 45 ms and 28 ms for the standard and richer cases at the same load. 
Our proposed SeLR method is significantly faster than the Optimal solver, taking 183 ms in the standard scenario with 100 tasks, and 132 ms, even when accounting for the cumulative scheduling time across multiple iterations as described in Algorithm~\ref{alg:task_offloading}.
These results demonstrate that for larger workloads involving 50–100 tasks in networks with 300 nodes, SeLR can accelerate the Optimal solver by 7.72--9.17$\times$ in standard scenarios, and up to 7.16--8.77$\times$ under richer server availability.
As shown in the experiments in Section~\ref{subsec:performancevsload}, Greedy-T$\times100$ improves upon Greedy-T and performs similarly to SeLR up to a task load of around 65. 
However, its computational overhead scales linearly with $M$, effectively running the Greedy-T approach $M$ times to select the best outcome.
This results in significantly higher computational costs for Greedy-T$\times100$ than the Optimal approach across all task loads. 
The overhead for Greedy-T$\times100$ under 50- and 100-task loads in both network settings reveals it is 2.52--2.63$\times$ slower than Optimal in standard scenarios and 2.9--2.29$\times$ slower in richer settings, as shown in Table~\ref{tab:overhead}.
A similar trend is observed for the Linear-Relax solver, previously noted in Section~\ref{subsec:performancevsload} for its high computational cost. It is 18--171.5$\times$ slower than Optimal in standard settings for 50–100 tasks and 10.65--44.64$\times$ slower in richer scenarios.
Due to these substantial overheads, we omit the scheduling time for Greedy-T$\times100$ and Linear-Relax from Fig.~\ref{fig:computationaltime} and report them separately in Table~\ref{tab:overhead}.

\section{Conclusions and Future Work}
In this study, we propose an iterative approach to transforming a non-convex problem formulation into a convex optimization for joint computation offloading and routing in edge networks.
Through numerical evaluations on simulated networks designed to closely resemble real-world datasets, our approach is demonstrated to achieve smaller optimality gap and better scalability compared to popular greedy heuristics with a comparable computational overhead, which is significantly lower than that of the optimal solver.  
Future directions include augmenting the scalability and optimality of our algorithmic framework with graph-based machine learning \cite{wu2020comprehensive, scarselli2008graph, zhou2020graph, xu2018powerful, chien2024opportunities, wolfe2024gist, isufi2024graph}.
In particular, instead of assigning all the tasks with integer solutions during the iterations, we may selectively withhold tasks from assignment by leveraging Graph Attention Networks (GATs) \cite{velivckovic2017graph} to predict their influence on unassigned tasks under residual resources to further reduce the optimality gap. 
% By learning impact scores through attention mechanisms, our method can make more informed task placement decisions, further improving efficiency and scalability in edge networks.

\bibliographystyle{ACM-Reference-Format}
\bibliography{SeLR}


\begin{thebibliography}{44}

%%% the Web address.

\ifx \showCODEN    \undefined \def \showCODEN     #1{\unskip}     \fi
\ifx \showISBNx    \undefined \def \showISBNx     #1{\unskip}     \fi
\ifx \showISBNxiii \undefined \def \showISBNxiii  #1{\unskip}     \fi
\ifx \showISSN     \undefined \def \showISSN      #1{\unskip}     \fi
\ifx \showLCCN     \undefined \def \showLCCN      #1{\unskip}     \fi
\ifx \shownote     \undefined \def \shownote      #1{#1}          \fi
\ifx \showarticletitle \undefined \def \showarticletitle #1{#1}   \fi
\ifx \showURL      \undefined \def \showURL       {\relax}        \fi
% invisible to TeX
\providecommand\bibfield[2]{#2}
\providecommand\bibinfo[2]{#2}
\providecommand\natexlab[1]{#1}
\providecommand\showeprint[2][]{arXiv:#2}

\bibitem[Belotti et~al\mbox{.}(2013)]%
        {belotti2013mixed}
\bibfield{author}{\bibinfo{person}{Pietro Belotti}, \bibinfo{person}{Christian Kirches}, \bibinfo{person}{Sven Leyffer}, \bibinfo{person}{Jeff Linderoth}, \bibinfo{person}{James Luedtke}, {and} \bibinfo{person}{Ashutosh Mahajan}.} \bibinfo{year}{2013}\natexlab{}.
\newblock \showarticletitle{Mixed-integer nonlinear optimization}.
\newblock \bibinfo{journal}{\emph{Acta Numerica}}  \bibinfo{volume}{22} (\bibinfo{year}{2013}), \bibinfo{pages}{1--131}.
\newblock


\bibitem[Bi et~al\mbox{.}(2021)]%
        {Bi2021Lyapunov}
\bibfield{author}{\bibinfo{person}{Suzhi Bi}, \bibinfo{person}{Liang Huang}, \bibinfo{person}{Hui Wang}, {and} \bibinfo{person}{Ying-Jun~Angela Zhang}.} \bibinfo{year}{2021}\natexlab{}.
\newblock \showarticletitle{Lyapunov-Guided Deep Reinforcement Learning for Stable Online Computation Offloading in Mobile-Edge Computing Networks}.
\newblock \bibinfo{journal}{\emph{IEEE Trans. Wireless Commun.}} \bibinfo{volume}{20}, \bibinfo{number}{11} (\bibinfo{year}{2021}), \bibinfo{pages}{7519--7537}.
\newblock
\href{https://doi.org/10.1109/TWC.2021.3085319}{doi:\nolinkurl{10.1109/TWC.2021.3085319}}


\bibitem[Bi et~al\mbox{.}(2020)]%
        {bi2020joint}
\bibfield{author}{\bibinfo{person}{Suzhi Bi}, \bibinfo{person}{Liang Huang}, {and} \bibinfo{person}{Ying-Jun~Angela Zhang}.} \bibinfo{year}{2020}\natexlab{}.
\newblock \showarticletitle{Joint optimization of service caching placement and computation offloading in mobile edge computing systems}.
\newblock \bibinfo{journal}{\emph{IEEE Trans. Wireless Commun.}} \bibinfo{volume}{19}, \bibinfo{number}{7} (\bibinfo{year}{2020}), \bibinfo{pages}{4947--4963}.
\newblock


\bibitem[Burer and Letchford(2012)]%
        {burer2012non}
\bibfield{author}{\bibinfo{person}{Samuel Burer} {and} \bibinfo{person}{Adam~N Letchford}.} \bibinfo{year}{2012}\natexlab{}.
\newblock \showarticletitle{Non-convex mixed-integer nonlinear programming: A survey}.
\newblock \bibinfo{journal}{\emph{Surveys in Operations Research and Management Science}} \bibinfo{volume}{17}, \bibinfo{number}{2} (\bibinfo{year}{2012}), \bibinfo{pages}{97--106}.
\newblock


\bibitem[Candes et~al\mbox{.}(2008)]%
        {candes2008enhancing}
\bibfield{author}{\bibinfo{person}{Emmanuel~J Candes}, \bibinfo{person}{Michael~B Wakin}, {and} \bibinfo{person}{Stephen~P Boyd}.} \bibinfo{year}{2008}\natexlab{}.
\newblock \showarticletitle{Enhancing sparsity by reweighted l1 minimization}.
\newblock \bibinfo{journal}{\emph{J. of Fourier Analysis and Apps.}}  \bibinfo{volume}{14} (\bibinfo{year}{2008}), \bibinfo{pages}{877--905}.
\newblock


\bibitem[Chien et~al\mbox{.}(2024)]%
        {chien2024opportunities}
\bibfield{author}{\bibinfo{person}{Eli Chien}, \bibinfo{person}{Mufei Li}, \bibinfo{person}{Anthony Aportela}, \bibinfo{person}{Kerr Ding}, \bibinfo{person}{Shuyi Jia}, \bibinfo{person}{Supriyo Maji}, \bibinfo{person}{Zhongyuan Zhao}, \bibinfo{person}{Javier Duarte}, \bibinfo{person}{Victor Fung}, \bibinfo{person}{Cong Hao}, {et~al\mbox{.}}} \bibinfo{year}{2024}\natexlab{}.
\newblock \showarticletitle{Opportunities and challenges of graph neural networks in electrical engineering}.
\newblock \bibinfo{journal}{\emph{{Nature Reviews Elec. Eng.}}} \bibinfo{volume}{1}, \bibinfo{number}{8} (\bibinfo{year}{2024}), \bibinfo{pages}{529--546}.
\newblock


\bibitem[Dinh et~al\mbox{.}(2017)]%
        {dinh2017offloading}
\bibfield{author}{\bibinfo{person}{Thinh~Quang Dinh}, \bibinfo{person}{Jianhua Tang}, \bibinfo{person}{Quang~Duy La}, {and} \bibinfo{person}{Tony~QS Quek}.} \bibinfo{year}{2017}\natexlab{}.
\newblock \showarticletitle{Offloading in mobile edge computing: Task allocation and computational frequency scaling}.
\newblock \bibinfo{journal}{\emph{IEEE Trans. Commun.}} \bibinfo{volume}{65}, \bibinfo{number}{8} (\bibinfo{year}{2017}), \bibinfo{pages}{3571--3584}.
\newblock


\bibitem[Erfaniantaghvayi et~al\mbox{.}(2024)]%
        {erfaniantaghvayi2024ant}
\bibfield{author}{\bibinfo{person}{Negar Erfaniantaghvayi}, \bibinfo{person}{Zhongyuan Zhao}, \bibinfo{person}{Kevin Chan}, \bibinfo{person}{Gunjan Verma}, \bibinfo{person}{Ananthram Swami}, {and} \bibinfo{person}{Santiago Segarra}.} \bibinfo{year}{2024}\natexlab{}.
\newblock \showarticletitle{Ant Backpressure Routing for Wireless Multi-hop Networks with Mixed Traffic Patterns}. In \bibinfo{booktitle}{\emph{MILCOM 2024-2024 IEEE Military Comms. Conf. (MILCOM)}}. IEEE, \bibinfo{pages}{1174--1179}.
\newblock


\bibitem[Floudas(1995)]%
        {floudas1995nonlinear}
\bibfield{author}{\bibinfo{person}{Christodoulos~A Floudas}.} \bibinfo{year}{1995}\natexlab{}.
\newblock \bibinfo{booktitle}{\emph{Nonlinear and mixed-integer optimization: fundamentals and applications}}.
\newblock \bibinfo{publisher}{Oxford University Press}.
\newblock


\bibitem[Gerogiannis et~al\mbox{.}(2022)]%
        {gerogiannis2022deep}
\bibfield{author}{\bibinfo{person}{Gerasimos Gerogiannis}, \bibinfo{person}{Michael Birbas}, \bibinfo{person}{Aimilios Leftheriotis}, \bibinfo{person}{Eleftherios Mylonas}, \bibinfo{person}{Nikolaos Tzanis}, {and} \bibinfo{person}{Alexios Birbas}.} \bibinfo{year}{2022}\natexlab{}.
\newblock \showarticletitle{Deep reinforcement learning acceleration for real-time edge computing mixed integer programming problems}.
\newblock \bibinfo{journal}{\emph{IEEE Access}}  \bibinfo{volume}{10} (\bibinfo{year}{2022}), \bibinfo{pages}{18526--18543}.
\newblock


\bibitem[{Google}(2023)]%
        {ortools}
\bibfield{author}{\bibinfo{person}{{Google}}.} \bibinfo{year}{2023}\natexlab{}.
\newblock \bibinfo{title}{{OR-Tools}}.
\newblock \bibinfo{howpublished}{\url{https://developers.google.com/optimization}}.
\newblock
\newblock
\shownote{Accessed: 2025-04-05}.


\bibitem[Guo et~al\mbox{.}(2023)]%
        {guo2023intelligent}
\bibfield{author}{\bibinfo{person}{Hongzhi Guo}, \bibinfo{person}{Xiaoyi Zhou}, \bibinfo{person}{Jiadai Wang}, \bibinfo{person}{Jiajia Liu}, {and} \bibinfo{person}{Abderrahim Benslimane}.} \bibinfo{year}{2023}\natexlab{}.
\newblock \showarticletitle{Intelligent task offloading and resource allocation in digital twin based aerial computing networks}.
\newblock \bibinfo{journal}{\emph{IEEE J. on Selected Areas in Comms.}} (\bibinfo{year}{2023}).
\newblock


\bibitem[Isufi et~al\mbox{.}(2024)]%
        {isufi2024graph}
\bibfield{author}{\bibinfo{person}{Elvin Isufi}, \bibinfo{person}{Fernando Gama}, \bibinfo{person}{David~I Shuman}, {and} \bibinfo{person}{Santiago Segarra}.} \bibinfo{year}{2024}\natexlab{}.
\newblock \showarticletitle{Graph filters for signal processing and machine learning on graphs}.
\newblock \bibinfo{journal}{\emph{IEEE Trans. Signal Process.}}  \bibinfo{volume}{72} (\bibinfo{year}{2024}), \bibinfo{pages}{4745--4781}.
\newblock


\bibitem[Jiang et~al\mbox{.}(2022)]%
        {jiang2022joint}
\bibfield{author}{\bibinfo{person}{Hongbo Jiang}, \bibinfo{person}{Xingxia Dai}, \bibinfo{person}{Zhu Xiao}, {and} \bibinfo{person}{Arun Iyengar}.} \bibinfo{year}{2022}\natexlab{}.
\newblock \showarticletitle{Joint task offloading and resource allocation for energy-constrained mobile edge computing}.
\newblock \bibinfo{journal}{\emph{IEEE Trans. Mobile Computing}} \bibinfo{volume}{22}, \bibinfo{number}{7} (\bibinfo{year}{2022}), \bibinfo{pages}{4000--4015}.
\newblock


\bibitem[Jiang et~al\mbox{.}(2023)]%
        {jiang2023joint}
\bibfield{author}{\bibinfo{person}{Hongbo Jiang}, \bibinfo{person}{Xingxia Dai}, \bibinfo{person}{Zhu Xiao}, {and} \bibinfo{person}{Arun Iyengar}.} \bibinfo{year}{2023}\natexlab{}.
\newblock \showarticletitle{Joint Task Offloading and Resource Allocation for Energy-Constrained Mobile Edge Computing}.
\newblock \bibinfo{journal}{\emph{IEEE Trans. Mobile Computing.}} \bibinfo{volume}{22}, \bibinfo{number}{7} (\bibinfo{year}{2023}), \bibinfo{pages}{4000--4015}.
\newblock
\href{https://doi.org/10.1109/TMC.2022.3150432}{doi:\nolinkurl{10.1109/TMC.2022.3150432}}


\bibitem[Kamran et~al\mbox{.}(2022)]%
        {Kamran2022deco}
\bibfield{author}{\bibinfo{person}{Khashayar Kamran}, \bibinfo{person}{Edmund Yeh}, {and} \bibinfo{person}{Qian Ma}.} \bibinfo{year}{2022}\natexlab{}.
\newblock \showarticletitle{{DECO}: Joint Computation Scheduling, Caching, and Communication in Data-Intensive Computing Networks}.
\newblock \bibinfo{journal}{\emph{IEEE/ACM Trans. Netw.}} \bibinfo{volume}{30}, \bibinfo{number}{3} (\bibinfo{year}{2022}), \bibinfo{pages}{1058--1072}.
\newblock
\href{https://doi.org/10.1109/TNET.2021.3136157}{doi:\nolinkurl{10.1109/TNET.2021.3136157}}


\bibitem[Kan et~al\mbox{.}(2018)]%
        {kan2018task}
\bibfield{author}{\bibinfo{person}{Te-Yi Kan}, \bibinfo{person}{Yao Chiang}, {and} \bibinfo{person}{Hung-Yu Wei}.} \bibinfo{year}{2018}\natexlab{}.
\newblock \showarticletitle{Task offloading and resource allocation in mobile-edge computing system}. In \bibinfo{booktitle}{\emph{2018 27th Wireless and Optical Comms. Conf. (WOCC)}}. IEEE, \bibinfo{pages}{1--4}.
\newblock


\bibitem[Kiamari and Krishnamachari(2022)]%
        {kiamari2022gcnscheduler}
\bibfield{author}{\bibinfo{person}{Mehrdad Kiamari} {and} \bibinfo{person}{Bhaskar Krishnamachari}.} \bibinfo{year}{2022}\natexlab{}.
\newblock \showarticletitle{{GCN}scheduler: Scheduling distributed computing applications using graph convolutional networks}. In \bibinfo{booktitle}{\emph{Proc. 1st Intl. Workshop on Graph Neural Netw.}} \bibinfo{pages}{13--17}.
\newblock


\bibitem[Kochenderfer(2019)]%
        {kochenderfer2019algorithms}
\bibfield{author}{\bibinfo{person}{MJ Kochenderfer}.} \bibinfo{year}{2019}\natexlab{}.
\newblock \bibinfo{booktitle}{\emph{Algorithms for Optimization}}.
\newblock \bibinfo{publisher}{The MIT Press, Cambridge}.
\newblock


\bibitem[Lee and Leyffer(2011)]%
        {lee2011mixed}
\bibfield{author}{\bibinfo{person}{Jon Lee} {and} \bibinfo{person}{Sven Leyffer}.} \bibinfo{year}{2011}\natexlab{}.
\newblock \bibinfo{booktitle}{\emph{Mixed integer nonlinear programming}}. Vol.~\bibinfo{volume}{154}.
\newblock \bibinfo{publisher}{Springer Science \& Business Media}.
\newblock


\bibitem[Lin et~al\mbox{.}(2020)]%
        {lin2020distributed}
\bibfield{author}{\bibinfo{person}{Rongping Lin}, \bibinfo{person}{Zhijie Zhou}, \bibinfo{person}{Shan Luo}, \bibinfo{person}{Yong Xiao}, \bibinfo{person}{Xiong Wang}, \bibinfo{person}{Sheng Wang}, {and} \bibinfo{person}{Moshe Zukerman}.} \bibinfo{year}{2020}\natexlab{}.
\newblock \showarticletitle{Distributed Optimization for Computation Offloading in Edge Computing}.
\newblock \bibinfo{journal}{\emph{IEEE Trans. Wireless Commun.}} \bibinfo{volume}{19}, \bibinfo{number}{12} (\bibinfo{year}{2020}), \bibinfo{pages}{8179--8194}.
\newblock
\href{https://doi.org/10.1109/TWC.2020.3019805}{doi:\nolinkurl{10.1109/TWC.2020.3019805}}


\bibitem[Liu et~al\mbox{.}(2020)]%
        {liu2020distributed}
\bibfield{author}{\bibinfo{person}{Boxi Liu}, \bibinfo{person}{Yang Cao}, \bibinfo{person}{Yue Zhang}, {and} \bibinfo{person}{Tao Jiang}.} \bibinfo{year}{2020}\natexlab{}.
\newblock \showarticletitle{A Distributed Framework for Task Offloading in Edge Computing Networks of Arbitrary Topology}.
\newblock \bibinfo{journal}{\emph{IEEE Trans. Wireless Commun.}} \bibinfo{volume}{19}, \bibinfo{number}{4} (\bibinfo{year}{2020}), \bibinfo{pages}{2855--2867}.
\newblock
\href{https://doi.org/10.1109/TWC.2020.2968527}{doi:\nolinkurl{10.1109/TWC.2020.2968527}}


\bibitem[Liu et~al\mbox{.}(2019)]%
        {liu2019dynamic}
\bibfield{author}{\bibinfo{person}{Chen-Feng Liu}, \bibinfo{person}{Mehdi Bennis}, \bibinfo{person}{Mérouane Debbah}, {and} \bibinfo{person}{H.~Vincent Poor}.} \bibinfo{year}{2019}\natexlab{}.
\newblock \showarticletitle{Dynamic Task Offloading and Resource Allocation for Ultra-Reliable Low-Latency Edge Computing}.
\newblock \bibinfo{journal}{\emph{IEEE Trans. Commun.}} \bibinfo{volume}{67}, \bibinfo{number}{6} (\bibinfo{year}{2019}), \bibinfo{pages}{4132--4150}.
\newblock
\href{https://doi.org/10.1109/TCOMM.2019.2898573}{doi:\nolinkurl{10.1109/TCOMM.2019.2898573}}


\bibitem[Müller et~al\mbox{.}(2015)]%
        {Müller2015computation}
\bibfield{author}{\bibinfo{person}{Sabrina Müller}, \bibinfo{person}{Hussein Al-Shatri}, \bibinfo{person}{Matthias Wichtlhuber}, \bibinfo{person}{David Hausheer}, {and} \bibinfo{person}{Anja Klein}.} \bibinfo{year}{2015}\natexlab{}.
\newblock \showarticletitle{Computation offloading in wireless multi-hop networks: Energy Minimization via multi-dimensional knapsack problem}. In \bibinfo{booktitle}{\emph{IEEE Annual Intl. Symp. on Personal, Indoor, and Mobile Radio Communications (PIMRC)}}. \bibinfo{pages}{1717--1722}.
\newblock
\href{https://doi.org/10.1109/PIMRC.2015.7343576}{doi:\nolinkurl{10.1109/PIMRC.2015.7343576}}


\bibitem[Ning et~al\mbox{.}(2020)]%
        {ning2020intelligent}
\bibfield{author}{\bibinfo{person}{Zhaolong Ning}, \bibinfo{person}{Kaiyuan Zhang}, \bibinfo{person}{Xiaojie Wang}, \bibinfo{person}{Lei Guo}, \bibinfo{person}{Xiping Hu}, \bibinfo{person}{Jun Huang}, \bibinfo{person}{Bin Hu}, {and} \bibinfo{person}{Ricky~YK Kwok}.} \bibinfo{year}{2020}\natexlab{}.
\newblock \showarticletitle{Intelligent edge computing in internet of vehicles: A joint computation offloading and caching solution}.
\newblock \bibinfo{journal}{\emph{{IEEE Trans. on Intl. Trans. Systems}}} \bibinfo{volume}{22}, \bibinfo{number}{4} (\bibinfo{year}{2020}), \bibinfo{pages}{2212--2225}.
\newblock


\bibitem[Patsias et~al\mbox{.}(2023)]%
        {patsias2023task}
\bibfield{author}{\bibinfo{person}{Vasilios Patsias}, \bibinfo{person}{Petros Amanatidis}, \bibinfo{person}{Dimitris Karampatzakis}, \bibinfo{person}{Thomas Lagkas}, \bibinfo{person}{Kalliopi Michalakopoulou}, {and} \bibinfo{person}{Alexandros Nikitas}.} \bibinfo{year}{2023}\natexlab{}.
\newblock \showarticletitle{Task allocation methods and optimization techniques in edge computing: A systematic review of the literature}.
\newblock \bibinfo{journal}{\emph{Future Internet}} \bibinfo{volume}{15}, \bibinfo{number}{8} (\bibinfo{year}{2023}), \bibinfo{pages}{254}.
\newblock


\bibitem[Scarselli et~al\mbox{.}(2008)]%
        {scarselli2008graph}
\bibfield{author}{\bibinfo{person}{Franco Scarselli}, \bibinfo{person}{Marco Gori}, \bibinfo{person}{Ah~Chung Tsoi}, \bibinfo{person}{Markus Hagenbuchner}, {and} \bibinfo{person}{Gabriele Monfardini}.} \bibinfo{year}{2008}\natexlab{}.
\newblock \showarticletitle{The graph neural network model}.
\newblock \bibinfo{journal}{\emph{IEEE Trans. on Neural Networks}} \bibinfo{volume}{20}, \bibinfo{number}{1} (\bibinfo{year}{2008}), \bibinfo{pages}{61--80}.
\newblock


\bibitem[Tawarmalani and Sahinidis(2013)]%
        {tawarmalani2013convexification}
\bibfield{author}{\bibinfo{person}{Mohit Tawarmalani} {and} \bibinfo{person}{Nikolaos~V Sahinidis}.} \bibinfo{year}{2013}\natexlab{}.
\newblock \bibinfo{booktitle}{\emph{Convexification and global optimization in continuous and mixed-integer nonlinear programming: theory, algorithms, software, and applications}}. Vol.~\bibinfo{volume}{65}.
\newblock \bibinfo{publisher}{Springer Science \& Business Media}.
\newblock


\bibitem[Tran and Pompili(2018)]%
        {tran2018joint}
\bibfield{author}{\bibinfo{person}{Tuyen~X Tran} {and} \bibinfo{person}{Dario Pompili}.} \bibinfo{year}{2018}\natexlab{}.
\newblock \showarticletitle{Joint task offloading and resource allocation for multi-server mobile-edge computing networks}.
\newblock \bibinfo{journal}{\emph{IEEE Trans. Vehicular Tech.}} \bibinfo{volume}{68}, \bibinfo{number}{1} (\bibinfo{year}{2018}), \bibinfo{pages}{856--868}.
\newblock


\bibitem[Ullah et~al\mbox{.}(2023)]%
        {ullah2023optimizing}
\bibfield{author}{\bibinfo{person}{Ihsan Ullah}, \bibinfo{person}{Hyun-Kyo Lim}, \bibinfo{person}{Yeong-Jun Seok}, {and} \bibinfo{person}{Youn-Hee Han}.} \bibinfo{year}{2023}\natexlab{}.
\newblock \showarticletitle{Optimizing task offloading and resource allocation in edge-cloud networks: a DRL approach}.
\newblock \bibinfo{journal}{\emph{{J. of Cloud Computing}}} \bibinfo{volume}{12}, \bibinfo{number}{1} (\bibinfo{year}{2023}), \bibinfo{pages}{112}.
\newblock


\bibitem[Veli{\v{c}}kovi{\'c} et~al\mbox{.}(2017)]%
        {velivckovic2017graph}
\bibfield{author}{\bibinfo{person}{Petar Veli{\v{c}}kovi{\'c}}, \bibinfo{person}{Guillem Cucurull}, \bibinfo{person}{Arantxa Casanova}, \bibinfo{person}{Adriana Romero}, \bibinfo{person}{Pietro Lio}, {and} \bibinfo{person}{Yoshua Bengio}.} \bibinfo{year}{2017}\natexlab{}.
\newblock \showarticletitle{Graph attention networks}.
\newblock \bibinfo{journal}{\emph{arXiv preprint arXiv:1710.10903}} (\bibinfo{year}{2017}).
\newblock


\bibitem[Wang et~al\mbox{.}(2019)]%
        {wang2019edge}
\bibfield{author}{\bibinfo{person}{Shangguang Wang}, \bibinfo{person}{Yali Zhao}, \bibinfo{person}{Jinlinag Xu}, \bibinfo{person}{Jie Yuan}, {and} \bibinfo{person}{Ching-Hsien Hsu}.} \bibinfo{year}{2019}\natexlab{}.
\newblock \showarticletitle{Edge server placement in mobile edge computing}.
\newblock \bibinfo{journal}{\emph{{J. of Parallel and Dist. Computing.}}}  \bibinfo{volume}{127} (\bibinfo{year}{2019}), \bibinfo{pages}{160--168}.
\newblock


\bibitem[Wolfe et~al\mbox{.}(2024)]%
        {wolfe2024gist}
\bibfield{author}{\bibinfo{person}{Cameron~R Wolfe}, \bibinfo{person}{Jingkang Yang}, \bibinfo{person}{Fangshuo Liao}, \bibinfo{person}{Arindam Chowdhury}, \bibinfo{person}{Chen Dun}, \bibinfo{person}{Artun Bayer}, \bibinfo{person}{Santiago Segarra}, {and} \bibinfo{person}{Anastasios Kyrillidis}.} \bibinfo{year}{2024}\natexlab{}.
\newblock \showarticletitle{GIST: Distributed training for large-scale graph convolutional networks}.
\newblock \bibinfo{journal}{\emph{{J. of Applied and Computat. Topology}}} \bibinfo{volume}{8}, \bibinfo{number}{5} (\bibinfo{year}{2024}), \bibinfo{pages}{1363--1415}.
\newblock


\bibitem[Wu et~al\mbox{.}(2020)]%
        {wu2020comprehensive}
\bibfield{author}{\bibinfo{person}{Zonghan Wu}, \bibinfo{person}{Shirui Pan}, \bibinfo{person}{Fengwen Chen}, \bibinfo{person}{Guodong Long}, \bibinfo{person}{Chengqi Zhang}, {and} \bibinfo{person}{Philip~S Yu}.} \bibinfo{year}{2020}\natexlab{}.
\newblock \showarticletitle{A comprehensive survey on graph neural networks}.
\newblock \bibinfo{journal}{\emph{IEEE Trans. on Neural Networks and Learning Systems}} \bibinfo{volume}{32}, \bibinfo{number}{1} (\bibinfo{year}{2020}), \bibinfo{pages}{4--24}.
\newblock


\bibitem[Xiao et~al\mbox{.}(2020b)]%
        {xiao2020parameter}
\bibfield{author}{\bibinfo{person}{Jun Xiao}, \bibinfo{person}{You Situ}, \bibinfo{person}{Weideng Yuan}, {and} \bibinfo{person}{Xinyang Wang}.} \bibinfo{year}{2020}\natexlab{b}.
\newblock \showarticletitle{Parameter Identification Method Based on Mixed-Integer Quadratic Programming and Edge Computing in Power Internet of Things}.
\newblock \bibinfo{journal}{\emph{{Math. Probs in Eng.}}} \bibinfo{volume}{2020}, \bibinfo{number}{1} (\bibinfo{year}{2020}), \bibinfo{pages}{4053825}.
\newblock


\bibitem[Xiao et~al\mbox{.}(2020a)]%
        {xiao2020edgeabc}
\bibfield{author}{\bibinfo{person}{Kaile Xiao}, \bibinfo{person}{Zhipeng Gao}, \bibinfo{person}{Weisong Shi}, \bibinfo{person}{Xuesong Qiu}, \bibinfo{person}{Yang Yang}, {and} \bibinfo{person}{Lanlan Rui}.} \bibinfo{year}{2020}\natexlab{a}.
\newblock \showarticletitle{EdgeABC: An architecture for task offloading and resource allocation in the Internet of Things}.
\newblock \bibinfo{journal}{\emph{Future Generation Computer Systems}}  \bibinfo{volume}{107} (\bibinfo{year}{2020}), \bibinfo{pages}{498--508}.
\newblock


\bibitem[Xu et~al\mbox{.}(2018)]%
        {xu2018powerful}
\bibfield{author}{\bibinfo{person}{Keyulu Xu}, \bibinfo{person}{Weihua Hu}, \bibinfo{person}{Jure Leskovec}, {and} \bibinfo{person}{Stefanie Jegelka}.} \bibinfo{year}{2018}\natexlab{}.
\newblock \showarticletitle{How powerful are graph neural networks?}
\newblock \bibinfo{journal}{\emph{arXiv preprint arXiv:1810.00826}} (\bibinfo{year}{2018}).
\newblock


\bibitem[Yan et~al\mbox{.}(2019)]%
        {yan2019optimal}
\bibfield{author}{\bibinfo{person}{Jia Yan}, \bibinfo{person}{Suzhi Bi}, \bibinfo{person}{Ying~Jun Zhang}, {and} \bibinfo{person}{Meixia Tao}.} \bibinfo{year}{2019}\natexlab{}.
\newblock \showarticletitle{Optimal task offloading and resource allocation in mobile-edge computing with inter-user task dependency}.
\newblock \bibinfo{journal}{\emph{IEEE Trans. Wireless Commun.}} \bibinfo{volume}{19}, \bibinfo{number}{1} (\bibinfo{year}{2019}), \bibinfo{pages}{235--250}.
\newblock


\bibitem[Yang and Zhong(2023)]%
        {yang2023task}
\bibfield{author}{\bibinfo{person}{Ziyan Yang} {and} \bibinfo{person}{Shaochun Zhong}.} \bibinfo{year}{2023}\natexlab{}.
\newblock \showarticletitle{Task offloading and resource allocation for edge-enabled mobile learning}.
\newblock \bibinfo{journal}{\emph{China Communications}} \bibinfo{volume}{20}, \bibinfo{number}{4} (\bibinfo{year}{2023}), \bibinfo{pages}{326--339}.
\newblock


\bibitem[Younis et~al\mbox{.}(2024)]%
        {younis2024energy}
\bibfield{author}{\bibinfo{person}{Ayman Younis}, \bibinfo{person}{Sumit Maheshwari}, {and} \bibinfo{person}{Dario Pompili}.} \bibinfo{year}{2024}\natexlab{}.
\newblock \showarticletitle{Energy-latency computation offloading and approximate computing in mobile-edge computing networks}.
\newblock \bibinfo{journal}{\emph{IEEE Transactions on Network and Service Management}} \bibinfo{volume}{21}, \bibinfo{number}{3} (\bibinfo{year}{2024}), \bibinfo{pages}{3401--3415}.
\newblock


\bibitem[Zhao et~al\mbox{.}(2020)]%
        {zhao2020improving}
\bibfield{author}{\bibinfo{person}{Xiaobo Zhao}, \bibinfo{person}{Minoo Hosseinzadeh}, \bibinfo{person}{Nathaniel Hudson}, \bibinfo{person}{Hana Khamfroush}, {and} \bibinfo{person}{Daniel~E Lucani}.} \bibinfo{year}{2020}\natexlab{}.
\newblock \showarticletitle{Improving the accuracy-latency trade-off of edge-cloud computation offloading for deep learning services}. In \bibinfo{booktitle}{\emph{2020 IEEE Globecom Workshops (GC Wkshps}}. IEEE, \bibinfo{pages}{1--6}.
\newblock


\bibitem[Zhao et~al\mbox{.}(2025)]%
        {zhao2025icassp}
\bibfield{author}{\bibinfo{person}{Z. Zhao}, \bibinfo{person}{J. Perazzone}, \bibinfo{person}{G. Verma}, \bibinfo{person}{K. Chan}, \bibinfo{person}{A. Swami}, {and} \bibinfo{person}{S. Segarra}.} \bibinfo{year}{2025}\natexlab{}.
\newblock \showarticletitle{Joint Task Offloading and Routing in Wireless Multi-hop Networks Using Biased Backpressure Algorithm}. In \bibinfo{booktitle}{\emph{IEEE Intl. Conf. on Acoustics, Speech and Signal Process. (ICASSP)}}. \bibinfo{pages}{1--5}.
\newblock


\bibitem[Zhao et~al\mbox{.}(2024)]%
        {zhao2024congestionaware}
\bibfield{author}{\bibinfo{person}{Zhongyuan Zhao}, \bibinfo{person}{Jake Perazzone}, \bibinfo{person}{Gunjan Verma}, {and} \bibinfo{person}{Santiago Segarra}.} \bibinfo{year}{2024}\natexlab{}.
\newblock \showarticletitle{Congestion-Aware Distributed Task Offloading in Wireless Multi-Hop Networks Using Graph Neural Networks}. In \bibinfo{booktitle}{\emph{IEEE Intl. Conf. on Acoustics, Speech and Signal Process. (ICASSP)}}. \bibinfo{pages}{8951--8955}.
\newblock
\href{https://doi.org/10.1109/ICASSP48485.2024.10447302}{doi:\nolinkurl{10.1109/ICASSP48485.2024.10447302}}


\bibitem[Zhou et~al\mbox{.}(2020)]%
        {zhou2020graph}
\bibfield{author}{\bibinfo{person}{Jie Zhou}, \bibinfo{person}{Ganqu Cui}, \bibinfo{person}{Shengding Hu}, \bibinfo{person}{Zhengyan Zhang}, \bibinfo{person}{Cheng Yang}, \bibinfo{person}{Zhiyuan Liu}, \bibinfo{person}{Lifeng Wang}, \bibinfo{person}{Changcheng Li}, {and} \bibinfo{person}{Maosong Sun}.} \bibinfo{year}{2020}\natexlab{}.
\newblock \showarticletitle{Graph neural networks: A review of methods and applications}.
\newblock \bibinfo{journal}{\emph{AI Open}}  \bibinfo{volume}{1} (\bibinfo{year}{2020}), \bibinfo{pages}{57--81}.
\newblock


\end{thebibliography}
\end{document}